\documentclass[a4paper, 12pt]{article}
\usepackage{graphicx}
\usepackage{amsmath}
\usepackage{theorem}

\title{Characterization Of A Class Of Graphs Related To Pairs Of Disjoint Matchings \thanks{The current work is the main part of the author's Master's thesis defended in May 2007.}}
\author{Anush Tserunyan}
\date{\small{Department of Informatics and Applied Mathematics, Yerevan State
University, \\ Yerevan, 0025, Armenia \\ Email: anush@math.ucla.edu,
anush\_tserunyan@yahoo.com \\} \scriptsize{The work on this paper
was supported by a grant of Armenian National Science and
Educational Fund.}}

\newtheorem{definition}{Definition}[section]
\newtheorem{property}[definition]{Property}
\newtheorem{lemmasection}[definition]{Lemma}
\newtheorem{corollarysection}[definition]{Corollary}
\newtheorem{lemma}{Lemma}[subsection]
\newtheorem{assumption}{Assumption}
\newtheorem{statement}{Statement}
\newtheorem{corollary}[lemma]{Corollary}

\theorembodyfont{\rmfamily}
\newtheorem{case}{Case}
\newtheorem{remark}{Remark}

\newenvironment{proof}[1][Proof]{\noindent\textbf{#1.} }{\ \rule{0.5em}{0.5em}}

\newenvironment{dedication}[1][]{\begin{trivlist}
\item[\hskip \labelsep {\bfseries #1}]}{\end{trivlist}}

\newenvironment{theorem}[1][Theorem]{\begin{trivlist}
\item[\hskip \labelsep {\bfseries #1}]}{\end{trivlist}}

\newenvironment{acknowledgements}[1][Acknowledgements]{\begin{trivlist}
\item[\hskip \labelsep {\bfseries #1}]}{\end{trivlist}}

\begin{document}
\maketitle

\begin{dedication}
$$\textit{Dedicated to my mother, father and sister Arevik}$$
\end{dedication}

\begin{abstract}
For a given graph consider a pair of disjoint matchings the union of
which contains as many edges as possible. Furthermore, consider the
ratio of the cardinalities of a maximum matching and the largest
matching in those pairs. It is known that for any graph
$\frac{5}{4}$ is the tight upper bound for this ratio. We
characterize the class of graphs for which it is precisely
$\frac{5}{4}$. Our characterization implies that these graphs
contain a spanning subgraph, every connected component of which is
the minimal graph of this class.
\end{abstract}

\section{Introduction}
In this paper we consider finite undirected graphs without multiple
edges, loops, or isolated vertices. Let $V(G)$ and $E(G)$ be the
sets of vertices and edges of a graph $G$, respectively.

We denote by $\beta(G)$ the cardinality of a maximum matching of
$G$.

Let $B_2(G)$ be the set of pairs of disjoint matchings of $G$. Set:
$$\lambda(G) \doteq max \{ |H|+|H'| : (H,H') \in B_2(G)\}.$$

Furthermore, let us introduce another parameter:
$$\alpha(G) \doteq max \{ |H|,|H'| : (H,H') \in B_2(G) \text{ and } |H|+|H'|=\lambda(G) \},$$
and define a set:
$$M_2(G) \doteq \{(H,H') \in B_2(G) : |H|+|H'|=\lambda(G) \text{ and } |H| = \alpha(G) \}.$$

While working on the problems of constructing a maximum matching $F$
of a graph $G$ such that $\beta(G \backslash F)$ is maximized or
minimized, Kamalian and Mkrtchyan designed polynomial algorithms for
solving these problems for trees \cite{alg}. Unfortunately, the
problems turned out to be NP-hard already for connected bipartite
graphs with maximum degree three \cite{np}, thus there is no hope
for the polynomial time calculation of $\beta_1(G)$ even for
bipartite graphs $G$, where $$\beta_1(G) = max \{ \beta(G \backslash
F) : \text{$F$ is a maximum matching of $G$} \}.$$

Note that for any graph $G$ $$\lambda(G) = \beta(G) + \beta_1(G)
\text{ if and only if } \alpha(G) = \beta(G).$$ Thus, $\beta_1(G)$
can be efficiently calculated for bipartite graphs $G$ with
$\beta(G) = \alpha(G)$ since $\lambda(G)$ can be calculated for that
graphs by using a standard algorithm of finding a maximum flow in a
network. Let us also note that the calculation of $\lambda(G)$ is
NP-hard even for the class of cubic graphs since the chromatic class
of a cubic graph $G$ is three if and only if $\lambda(G) = |V(G)|$
(see \cite{Holyer}).

Being interested in the classification of graphs $G$, for which
$\beta(G) = \alpha(G)$, Mkrtchyan in \cite{MPP01} proved a
sufficient condition, which due to \cite{Har, Har-Plum}, can be
formulated as: if $G$ is a matching covered tree then $\beta(G) =
\alpha(G)$. Note that a graph is said to be matching covered (see
\cite{Perfect}) if its every edge belongs to a maximum matching (not
necessarily a perfect matching as it is usually defined, see e.g.
\cite{Lov-Plum}).

In contrast with the theory of 2-matchings, where every graph $G$
admits a maximum 2-matching that includes a maximum matching
\cite{Lov-Plum}, there are graphs (even trees) that do not have a
``maximum" pair of disjoint matchings (a pair from $M_2(G)$) that
includes a maximum matching.

The following is the best result that can be stated about the ratio
$\frac{\beta(G)}{\alpha(G)}$ for any graph $G$ (see \cite{VAV}):
$$1 \leq \frac{\beta(G)}{\alpha(G)} \leq \frac{5}{4}.$$

The aim of the paper is the characterization of the class of graphs
$G$, for which the ratio $\frac{\beta(G)}{\alpha(G)}$ obtains its
upper bound, i.e. the equality $\frac{\beta(G)}{\alpha(G)} =
\frac{5}{4}$ holds.

\begin{figure}[h]
\begin{center}
\includegraphics{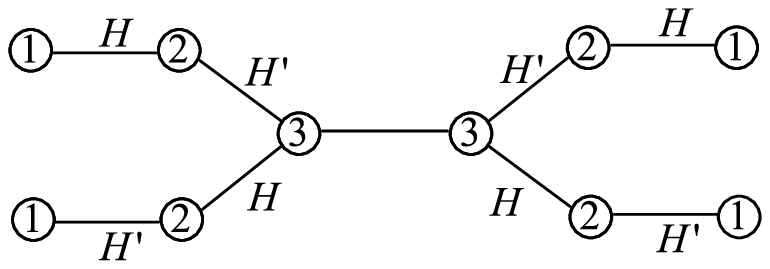}\\
\caption{Spanner}\label{fig_spanner}
\end{center}
\end{figure}

Our characterization theorem is formulated in terms of a special
graph called spanner (figure \ref{fig_spanner}), which is the
minimal graph for which $\beta \neq \alpha$ (what is remarkable is
that the equality $\frac{\beta}{\alpha} = \frac{5}{4}$ also holds
for spanner). This kind of theorems is common in graph theory: see
\cite{Har} for characterization of planar or line graphs. Another
example may be Tutte's Conjecture (now a beautiful theorem thanks to
Robertson, Sanders, Seymour and Tomas) about the chromatic index of
bridgeless cubic graphs, which do not contain Petersen graph as a
minor.

On the other hand, let us note that in contrast with the examples
given above, our theorem does not provide a forbidden/excluded graph
characterization. Quite the contrary, the theorem implies that every
graph satisfying the mentioned equality admits a spanning subgraph
every connected component of which is a spanner.

\section{Main Notations and Definitions}

Let $G$ be a graph and $d_G(v)$ be the degree of a vertex $v$ of
$G$.

\begin{definition}\label{matching_definition}
A subset of $E(G)$ is called a matching if it does not contain
adjacent edges.
\end{definition}

\begin{definition}\label{maximum_matching_definition}
A matching of $G$ with maximum number of edges is called maximum.
\end{definition}

\begin{definition}\label{covered_missed_by_matching}
A vertex $v$ of $G$ is covered (missed) by a matching $H$ of $G$, if
$H$ contains (does not contain) an edge incident to $v$.
\end{definition}

\begin{definition}\label{trail_definition}
A sequence $v_0, e_1, v_1, ..., v_{n-1}, e_n, v_n$ is called a trail
in $G$ if $v_i \in V(G)$, $e_j \in E(G)$, $e_j = (v_{j-1}, v_j)$,
and $e_j \neq e_k$ if $j \neq k$, for $0 \leq i \leq n$, $1 \leq j,k
\leq n$.
\end{definition}

The number of edges, $n$, is called the length of a trail $v_0, e_1,
v_1, ..., v_{n-1}, e_n, v_n$. Trail is called even (odd) if its
length is even (odd).

Trails $v_0, e_1, v_1, ..., v_{n-1}, e_n, v_n$ and $v_n, e_n,
v_{n-1}, ..., v_1, e_1, v_0$ are considered equal. Trail $T$ is also
considered as a subgraph of $G$, and thus, $V(T)$ and $E(T)$ are
used to denote the sets of vertices and edges of $T$, respectively.

\begin{definition}\label{cycle_definition}
A trail $v_0, e_1, v_1, ..., v_{n-1}, e_n, v_n$ is called a cycle if
$v_0 = v_n$.
\end{definition}

Similarly, cycles $v_0, e_1, v_1, ..., v_{n-1}, e_n, v_0$ and $v_i,
e_{i+1}, ..., e_n, v_0, e_1, ..., e_i, v_i$ are considered equal for
any $0 \leq i \leq n-1$.

If $T : v_0, e_1, v_1, ..., v_{n-1}, e_n, v_n$ is a trail that is
not a cycle then $v_0$, $v_n$ and $e_1$, $e_n$ are called the
end-vertices and end-edges of $T$, respectively.

\begin{definition}\label{path_definition}
A trail $v_0, e_1, v_1, ..., v_{n-1}, e_n, v_n$ is called a path if
$v_i \neq v_j$ for $0 \leq i < j \leq n$.
\end{definition}

\begin{definition}\label{simple_cycle_definition}
A cycle $v_0, e_1, v_1, ..., v_{n-1}, e_n, v_0$ is called simple if
$v_0, e_1, v_1, ..., v_{n-2}, e_{n-1}, v_{n-1}$ is a path.
\end{definition}

Below we omit $v_i$-s and write $e_1, e_2, ..., e_n$ instead of
$v_0, e_1, v_1, ..., v_{n-1}, e_n, v_n$ when denoting a trail.

\begin{definition}\label{end_edges_and_vertices}
For a trail $T : e_1, e_2, ..., e_n$ of $G$ and $i \geq 1$, define
sets $E_i^b(T)$, $E_i^e(T)$, $E_i(T)$ and $V_0(T)$, $V_i^b(T)$,
$V_i^e(T)$, $V_i(T)$ as follows:
$$E_i^b(T) \doteq \{ e_j : 1 \leq j \leq min\{n, i\} \},$$
$$E_i^e(T) \doteq \{ e_j : max\{1, n-i+1\} \leq j \leq n \},$$
$$E_i(T) \doteq E_i^b(T) \cup E_i^e(T),$$
and
$$V_0(T) \doteq \{ v \in V(G) : v \text{ is an end-vertex of } T \},$$
$$V_i^b(T) \doteq \{ v \in V(G) : v \text{ is incident to an edge from } E_i^b(T) \},$$
$$V_i^e(T) \doteq \{ v \in V(G) : v \text{ is incident to an edge from } E_i^e(T) \},$$
$$V_i(T) \doteq V_i^b(T) \cup V_i^e(T).$$
\end{definition}

The same notations are used for sets of trails. For example, for a
set of trails $D$, $V_0(D)$ denotes the set of end-vertices of all
trails from $D$, that is:
$$V_0(D) \doteq \bigcup_{T \in D}V_0(T).$$

Let $A$ and $B$ be sets of edges of $G$.
\begin{definition}\label{alt_trail_definition}
A trail $e_1, e_2, ..., e_n$ is called $A$-$B$ alternating if the
edges with odd indices belong to $A \backslash B$ and others to $B
\backslash A$, or vice-versa.
\end{definition}

If $X$ is an $A$-$B$ alternating trail then $X_A$ ($X_B$) denotes
the graph induced by the set of edges of $X$ that belong to $A$
($B$).

The set of $A$-$B$ alternating trails of $G$ that are not cycles is
denoted by $T(A,B)$. The subsets of $T(A,B)$ containing only even
and odd trails are denoted by $T_e(A,B)$ and $T_o(A,B)$,
respectively. We use the notation $C$ instead of $T$ do denote the
corresponding sets of $A$-$B$ alternating cycles (e.g. $C_e(A,B)$ is
the set of $A$-$B$ alternating even cycles).

The set of the trails from $T_o(A,B)$ starting with an edge from $A$
($B$) is denoted by $T_o^A(A,B)$ ($T_o^B(A,B)$).

Now, let $A$ and $B$ be matchings of $G$ (not necessarily disjoint).
Note that $A$-$B$ alternating trail is either a path, or an even
simple cycle.

\begin{definition}\label{max_alt_path_definition}
An $A$-$B$ alternating path $P$ is called maximal if there is no
other $A$-$B$ alternating trail (a path or an even simple cycle)
that contains $P$ as a proper subtrail.
\end{definition}

We use the notation $MP$ instead of $T$ to denote the corresponding
sets of maximal  $A$-$B$ alternating paths (e.g. $MP_o^B(A,B)$ is
the subset of $MP(A,B)$ containing only those maximal $A$-$B$
alternating paths whose length is odd and which start (and also end)
with an edge from $B$).

Terms and concepts that we do not define can be found in \cite{Har,
Lov-Plum, West}.

\section{General Properties and Structural Lemmas}

Let $G$ be a graph, and $A$ and $B$ be (not necessarily disjoint)
matchings of it. The following are properties of $A$-$B$ alternating
cycles and maximal paths.

\bigskip

First note that all cycles from $C_e(A,B)$ are simple as $A$ and $B$
are matchings.

\begin{property}\label{path_evencycle_HH'}
If the connected components of $G$ are paths or even simple cycles,
and \\ $(H, H') \in M_2(G)$, then $H \cup H' = E(G)$.
\end{property}

\begin{property}\label{altpathequaldegrees}
If $C \in C_e(A,B)$ and $v\in V(C)$ then $d_{C_A}(v) = d_{C_B}(v)$.
\end{property}

\begin{property}\label{edge_lies_on_alt_component}
Every edge $e \in A \triangle B$ \footnote{$A \triangle B$ denotes
the symmetric difference of $A$ and $B$, i.e. $A \triangle B = (A
\backslash B) \cup (B \backslash A).$} lies either on a cycle from
$C_e(A,B)$ or on a path from $MP(A,B)$.
\end{property}

\begin{property}\label{AB}\
\renewcommand{\labelenumi}{(\arabic{enumi})}
\begin{enumerate}
\item if $F \in C_e(A,B) \cup T_e(A,B)$ then $A$ and $B$ have the same number of edges that
lie on $F$,

\item if $T \in T_o^A(A,B)$ then the number of edges from $A$ lying on $T$ is one more than the number of ones from $B$.
\end{enumerate}
\end{property}

These observations imply:

\begin{property}\label{cardinalitydiff}
$|A|-|B|=|MP_{o}^{A}(A,B)|-|MP_{o}^{B}(A,B)|$.
\end{property}

Berge's well-known theorem states that a matching $M$ of a graph $G$
is maximum if and only if $G$ does not contain an $M$-augmenting
path \cite{Har, Lov-Plum, West}. This theorem immediately implies:

\begin{property}
\label{maxmatchingproperty} If $M$ is a maximum matching and $H$ is
a matching of a graph $G$ then
\begin{equation*}
MP_{o}^{H}(M,H)=\emptyset,
\end{equation*}%
and therefore, $|M|-|H|=|MP_{o}^{M}(M,H)|$.
\end{property}

The proof of the following property is similar to the one of
property \ref{maxmatchingproperty}:

\begin{property} \label{HH'}
If $(H,H')\in M_2(G)$ then $MP_o^{H'}(H,H') = \emptyset $.
\end{property}

\begin{property} \label{lambda=2alpha}
If $\lambda(G) = 2\alpha(G)$ and $(H,H') \in B_2(G)$ for which $|H|
+ |H'| = \lambda(G)$, then $MP_o(H,H') = \emptyset$ and $(H,H') \in
M_2(G)$.
\end{property}
\begin{proof}
Assume that $MP_o(H,H') \neq \emptyset$. Denote by $O$ and $E$ the
sets of edges lying on the paths from $MP_o(H,H')$ with odd and even
indices, respectively (indices start with $1$). Set
$$H_1 = (H \backslash E) \cup O,$$
and
$$H_1' = (H' \backslash O) \cup E.$$
Note that $(H_1, H_1') \in B_2(G)$ and $|H_1| + |H_1'| = |H| + |H'|
= \lambda(G)$ (as $H_1 \cup H_1' = H \cup H'$). Also note that
$MP_o^{H_1}(H_1, H_1') = MP_o(H, H')$ and $MP_o^{H_1'}(H_1, H_1') =
\emptyset$. Due to property \ref{cardinalitydiff}, $|H_1| - |H_1'| =
|MP_o^{H_1}(H_1, H_1')| = |MP_o(H, H')| > 0$, i.e. $|H_1| > |H_1'|$,
and therefore $|H_1| > \frac{\lambda(G)}{2} = \alpha(G)$, which
contradicts the definition of $\alpha(G)$. Thus, $MP_o(H,H') =
\emptyset$, and, due to property \ref{cardinalitydiff}, $|H| = |H'|
= \frac{\lambda(G)}{2} = \alpha(G)$, which means that $(H,H') \in
M_2(G)$.
\end{proof}


\bigskip

Now let $M$ be a fixed maximum matching of $G$. Over all $(H,H')\in
M_{2}(G)$, consider the pairs $((H,H'),M)$ for which $|M \cap (H
\cup H')|$ is maximized. Denote the set of those pairs by $M_2(G,
M)$:
$$M_2(G, M) \doteq \{(H,H') \in M_{2}(G) : |M \cap (H \cup H')| \text{ is maximum} \}.$$

Let $(H,H')$ be an arbitrarily chosen pair from $M_2(G, M)$.

\begin{lemmasection}\label{MHmain_property}
For every path $P : m_1, h_1, ..., m_{l-1}, h_{l-1}, m_l$ from
$MP_o^M(M,H)$
\renewcommand{\labelenumi}{(\arabic{enumi})}
\begin{enumerate}
\item \label{endedgesH'} $m_1, m_l \in H'$;
\item \label{l>=3} $l\geq 3$.
\end{enumerate}
\end{lemmasection}
\begin{proof}
Let us show that $m_1, m_l \in H'$. If $l = 1$ then $P = m_1$, $m_1
\in M \backslash H$, and $m_1$ is not adjacent to an edge from $H$
as $P$ is maximal. Thus, $m_1 \in H'$ as otherwise we could enlarge
$H$ by adding $m_1$ to it which would contradict $(H, H') \in
M_2(G)$. Thus, suppose that $l \geq 2$. Let us show that $m_1 \in
H'$. If $m_1 \notin H'$ then define
$$H_1 \doteq (H \backslash \{h_1\}) \cup \{m_1\}.$$

Clearly, $H_1$ is a matching, and $H_{1}\cap H^{\prime }=\emptyset,
\ |H_1| = |H|,$ which means that $(H_1, H') \in M_2(G)$. But $|M
\cap (H_1 \cup H')| > |M \cap (H \cup H')|,$ which contradicts $(H,
H') \in M_2(G, M)$. Thus $m_1 \in H'$. Similarly, it can be shown
that $m_l \in H'$.

Now let us show that $l\geq 3$. Due to property \ref{HH'},
$MP_{o}^{H^{\prime }}(H,H^{\prime })=\emptyset $, thus there is $i,$
$1\leq i\leq l$, such that $m_i \in M \backslash (H \cup H')$, since
$\{m_1, m_l\} \subseteq H'$, and we have $l \geq 3$.
\end{proof}

\begin{lemmasection}\label{M_HandH's}
Each vertex lying on a path from $MP_o^M(M,H)$ is incident to an
edge from $H'$.
\end{lemmasection}
\begin{proof}
Assume the contrary, and let $v$ be a vertex lying on a path $P$
from $MP_o^M(M,H)$, which is not incident to an edge from $H'$.
Clearly, $v$ is incident to an edge $e = (u, v) \in M \backslash (H
\cup H')$ lying on $P$.

If $u$ is not incident to an edge from $H'$ too, then $H$ and $H'
\cup \{e\}$ are disjoint matchings and $|H|+|H' \cup
\{e\}|>|H|+|H'|=\lambda _{2}(G),$ which contradicts $(H,H') \in
M_2(G)$.

On the other hand, if $u$ is incident to an edge $f\in H'$, then
consider the pair $(H,H'')$, where $H'' \doteq (H' \backslash \{f\})
\cup \{e\}$. Note that $H$ and $H''$ are disjoint matchings and
$|H''| = |H'|$, which means that $(H, H'') \in M_2(G)$. But $|M \cap
(H \cup H'')| > |M \cap (H \cup H')|$ contradicting $(H, H') \in
M_2(G, M)$.
\end{proof}

For a path $P \in MP_o^M(M, H)$, consider one of its end-edges $f
\in E_1(P)$. Due to statement (\ref{endedgesH'}) of lemma
\ref{MHmain_property}, $f \in M \cap H'$. By maximality of $P$, $f$
is adjacent to only one edge from $H$, thus it is an end-edge of a
path $P_f$ from $MP_e(H,H') \cup MP_o^{H'}(H,H')$. Moreover, $P_f
\in MP_e(H,H')$ according to property \ref{HH'}. Define a set $Y
\subseteq MP_e(H,H')$ as follows: $$Y(M,H,H') \doteq \{P_f : P \in
MP_o^M(M, H), f \in E_1(P) \}.$$

\begin{lemmasection}\label{Y_stuff}\
\begin{enumerate}
\renewcommand{\labelenumi}{(\arabic{enumi})}
\item \label{evendisjoint} The end-edges of paths of $MP_o^M(M,H)$
lie on different paths of $Y(M,H,H')$;
\item \label{Yis2(b-a)} $|Y(M,H,H')| = 2|MP_o^M(M,H)| = 2(\beta(G) -
\alpha(G))$.
\item For every $P \in Y(M,H,H'), P : h'_1, h_1, ..., h'_n,
h_n$, where $h'_i \in H'$, $h_i \in H$, $1 \leq i \leq n$,
\begin{enumerate}
\renewcommand{\labelenumii}{(\alph{enumii})}
\item \label{povorot}
$h'_1$ and $h_1$ lie on a path from $MP_o^M(M,H)$, but $h'_n$ and
$h_n$ do not lie on any path from $MP_o^M(M,H)$;
\item \label{H'-H>=4} $n \geq 2$.
\end{enumerate}
\end{enumerate}
\end{lemmasection}
\begin{proof}
(\ref{evendisjoint}) is true as otherwise we would have a path from
$Y(M,H,H')$ with both end-edges from $H'$ contradicting $Y(M,H,H')
\subseteq MP_e(H,H')$. Furthermore, (\ref{evendisjoint}) together
with property \ref{maxmatchingproperty} imply (\ref{Yis2(b-a)}).

Now, let us prove (\ref{povorot}).

By the definition of $Y(M,H,H')$, $h'_1$ is an end-edge of a path
$P_1$ from $MP_o^M(M,H)$, and therefore $h_1$ lies on $P_1$ too.

$h_n$ does not lie on any path from $MP_o^M(M,H)$ as otherwise, due
to lemma \ref{M_HandH's}, both vertices incident to $h_n$ would be
incident to edges from $H'$, which contradicts the maximality of
$P$. Note that $h_n$ is not incident to an inner vertex (not an
end-vertex) of a path $P_1$ from $MP_o^M(M,H)$ as any such vertex is
incident to an edge from $H$ lying on $P_1$, and therefore different
from $h_n$. $h_n$ is incident neither to an end-vertex of a path
$P_1$ from $MP_o^M(M,H)$ as it would contradict the maximality of
$P_1$. Thus, $h_n$ is not adjacent to an edge lying on a path from
$MP_o^M(M,H)$, and therefore $h'_n$ does not lie on any path from
$MP_o^M(M,H)$. The proof of (\ref{povorot}) is complete.

Statement (\ref{H'-H>=4}) immediately follows from (\ref{povorot}).
\end{proof}

Taking into account that $|H| = \alpha(G)$, $|H'| = \lambda(G) -
\alpha(G)$, and $|H| \geq |H'|$, we get the following result (also
obtained in \cite{VAV}) as a corollary from the statements
(\ref{Yis2(b-a)}) and (\ref{H'-H>=4}) of lemma \ref{Y_stuff}:

\begin{corollarysection}\label{fivefourthinequality}
$\alpha(G) \geq \lambda(G) - \alpha(G) \geq 4(\beta(G) -
\alpha(G))$, i.e. $\frac{\beta(G)}{\alpha(G)} \leq \frac{5}{4}$.
\end{corollarysection}

\section{Spanner, S-Forest and S-Graph}

The graph on figure \ref{fig_spanner} is called spanner. A vertex
$v$ of spanner $S$ is called $i$-vertex, $1\leq i \leq 3$, if
$d_S(v)=i$. The $3$-vertex closest to a vertex $v$ of spanner is
referred as the base of $v$. The two paths of the spanner of length
four connecting $1$-vertices are called {\em sides}.

For spanner $S$ define sets $U(S)$ and $L(S)$ as follows:

$$U(S) \doteq \{e \in E(S) : \text {$e$ is incident to a
1-vertex} \},$$
$$L(S) \doteq \{e \in E(S) \backslash U(S) : \text{$e$ is incident to a 2-vertex}\}.$$

Note that for spanner $S$, and for every $(H,H')\in M_2(S)$, the
edge connecting the $3$-vertices does not belong to $H\cup H'$,
hence $\lambda(S)=8$, $\alpha(S)=\lambda(S)-\alpha(S)=4$, and
$$\frac{\beta(S)}{\alpha(S)}=\frac{5}{4},$$ as $\beta(S)=5$. The pair $(H,H')$ shown on figure \ref{fig_spanner} belongs to
$M_2(S)$.

It can be implied from the lemma \ref{Y_stuff} that spanner is the
minimal graph for which the parameters $\alpha$ and $\beta$ are not
equal.

\begin{property} \label{coveredbyboth}
For spanner $S$ and $(H,H')\in M_2(S)$, $2$-vertices and
$3$-vertices of $S$ are covered by both $H$ and $H'$.
\end{property}

\begin{property} \label{missedbyone}
For every $1$-vertex $v$ of spanner $S$ there is $(H,H')\in M_2(S)$
such that $v$ is missed by $H$ ($H'$).
\end{property}

\begin{definition}\label{SForest}
$S$-forest is a forest whose connected components are spanners.
\end{definition}

An $i$-vertex of a connected component (spanner) of an $S$-forest
$F$ is referred simply as an $i$-vertex of $F$.

If $S_1,S_2,...,S_k$ are connected components of $S$-forest $F$ then
define sets $U(F)$ and $L(F)$ as follows:
$$U(F) \doteq \bigcup_{i=1}^k{U(S_i)};$$
$$L(F) \doteq \bigcup_{i=1}^k{L(S_i)}.$$

\begin{property} \label{S-forest_lambda=2alpha=8k}
If the number of connected components (spanners) of an $S$-forest
$F$ is $k$, then $\lambda(F) = 2\alpha(F) = 8k$, and $\beta(F) =
5k$, thus $\frac{\beta(F)}{\alpha(F)} = \frac{5}{4}$.
\end{property}

\begin{property} \label{S-forest_HH'_is_LU}
If $F$ is an $S$-forest and $(H,H') \in M_2(F)$ then $H \cup H' =
U(F) \cup L(F)$.
\end{property}

\begin{definition}\label{SGraph}
$S$-graph is a graph containing an $S$-forest as a spanning
subgraph (below, we will refer to it as a spanning $S$-forest of an
$S$-graph).
\end{definition}

Note that, spanning $S$-forest of an $S$-graph is not unique in
general.

It is easy to see that spanner, $S$-forests, and $S$-graphs contain
a perfect matching, and for $S$-forest it is unique.

Let $G$ be an $S$-graph with a spanning $S$-forest $F$.

\begin{property} \label{S-graph_beta=5k}
If $F$ has $k$ connected components (spanners) then $\beta(G) =
\beta(F) = 5k$.
\end{property}

Let us define an $i$-$j$ edge of $F$ as an edge connecting an
$i$-vertex to a $j$-vertex of $F$. Also define:
$$\Delta(G,F) \doteq \{e \in E(G) : \text{$e$ connects a $1$-vertex of $F$ to its base}\},$$

$$B(G,F) \doteq E(G) \backslash (L(F)\cup U(F) \cup \Delta(G, F)).$$

\begin{property}\label{2_and_3_are_equal}
For any $L(F)$-$B(G,F)$ alternating even cycle the numbers of
$2$-$2$ and $3$-$3$ edges lying on it are equal.
\end{property}
\begin{proof}
Consider an $L(F)$-$B(G,F)$ alternating even cycle $$C : (u_1, v_1),
(v_1, u_2), (u_2, v_2), ..., (v_{n-1}, u_n), (u_n, v_n), (v_n,
u_1),$$ where $(u_i, v_i) \in L(F), i = 1, 2, ..., n;$ $(v_n, u_1)
\in B(G,F), (v_j, u_{j+1}) \in B(G,F), j = 1, 2, ..., n-1$.

For a vertex $w$ of the cycle $C$ let $\delta(w)$ be the frequency
of appearance of the vertex $w$ during the circumference of $C$ (the
number of indices $i_0$ for which $w = u_{i_0}$ or $w = v_{i_0}$).
As any vertex $w$ lying on $C$ is incident to an edge from $L(F)$
that lies on $C$ before or after $v$ during the circumference, and
as edges from $L(F)$ are $2$-$3$ edges, we get:

$$\sum_{w \text{ is a 2-vertex lying on } C}\delta(w) = \sum_{w \text{ is a 3-vertex lying on } C}\delta(w) = n.$$

On the other hand, denote by $m_{22}, m_{33}, m_{23}$ the numbers of
$2$-$2$, $3$-$3$, $2$-$3$ edges lying on $C$, respectively. As for
each vertex $w$ lying on $C$, $2\delta(w)$ is the number of edges
that lie on $C$ and are incident to $w$, implies:
$$\sum_{w \text{ is a 2-vertex lying on } C} 2 \delta(w) = 2m_{22} + m_{23},$$
$$\sum_{w \text{ is a 3-vertex lying on } C} 2 \delta(w) = 2m_{33} + m_{23},$$
where the left sides of the equalities represent the numbers of
edges lying on $C$ and incident to $2$-vertices and $3$-vertices of
$C$, respectively. Thus, $m_{22} = m_{33}$.
\end{proof}

\section{Main Theorem}

\begin{theorem}
\renewcommand{\labelenumi}{(\alph{enumi})}
\textit{ For a graph $G$ ($G$ does not contain isolated vertices),
the equality $\frac{\beta(G)}{\alpha(G)}=\frac{5}{4}$ holds, if and
only if $G$ is an $S$-graph with a spanning $S$-forest $F$,
satisfying the following conditions:
\begin{enumerate}
\item \label{one-vertexstuff} $1$-vertices of $F$ are not incident to any edge from $B(G,F)$;
\item \label{two-vertexstuff} if a $1$-vertex $u$ of $F$ is incident to an edge from $\Delta(G,F)$, then
the\footnote{We write ``the'' here as if the condition (a) is
satisfied then there is only one $2$-vertex adjacent to $u$ (the
$2$-vertex connected to $u$ via the edge from $U(F)$).} $2$-vertex
of $F$ adjacent to $u$ is not incident to any edge from $B(G,F)$;
\item \label{LBaltcycles} for every $L(F)$-$B(G,F)$ alternating even cycle $C$ of $G$ containing a $2$-$2$ edge,
the graph $C_{B(G,F)}$ is not bipartite.
\end{enumerate}
}
\end{theorem}

The proof of the theorem is long, so it is divided into subsections:
Necessity and Sufficiency, which, in their turn, are split into
numbers of lemmas.

\subsection{Necessity}

In this subsection, we assume that
$\frac{\beta(G)}{\alpha(G)}=\frac{5}{4}$, and prove that $G$ is an
$S$-graph. Then, on the contrary assumptions we prove consequently
that the conditions (a), (b) and (c) are satisfied for an arbitrary
spanning $S$-forest of $G$. As one can see, we prove a statement
stronger than the Necessity of the theorem.

\bigskip

Let $G$ be a graph, $M$ be a fixed maximum matching of it, and
$(H,H')$ be an arbitrarily chosen pair from $M_2(G, M)$.

Suppose that for the graph $G$ the equality
$\frac{\beta(G)}{\alpha(G)}=\frac{5}{4}$ holds.

Due to corollary \ref{fivefourthinequality}, we have:
\renewcommand{\theequation}{\fnsymbol{equation}}
\setcounter{equation}{2}
\begin{equation}\label{fivefourthequality}
\alpha(G) = \lambda(G) - \alpha(G) = 4(\beta(G) - \alpha(G)).
\end{equation}

\begin{lemma}\label{MH5_HH'4}
Each path from $Y(M,H,H')$ is of length four, each path from
$MP_o^M(M,H)$ is of length five, and every edge from $H$ lies on a
path from $Y(M,H,H')$.
\end{lemma}
\begin{proof}
Due to equality (\ref{fivefourthequality}) and statement
(\ref{Yis2(b-a)}) of lemma \ref{Y_stuff}, we get:
$$|H| = 2|Y(M,H,H')|.$$
Therefore, as there are at least two edges from $H$ lying on each
path of $Y(M,H,H')$ (statement (\ref{H'-H>=4}) of the lemma
\ref{Y_stuff}), the length of each path from $Y(M,H,H')$ is
precisely four, and every edge from $H$ lies on a path from
$Y(M,H,H')$. Moreover, the length of every path from $MP_o^M(M,H)$
is precisely five (due to statement (\ref{l>=3}) of the lemma
\ref{MHmain_property} it is at least five for any graph), as
otherwise we would have either an edge from $H$ not lying on any
path from $Y(M,H,H')$, or a path from $Y(M,H,H')$ with length
greater than four.
\end{proof}

\begin{figure}[h]
\begin{center}
\includegraphics{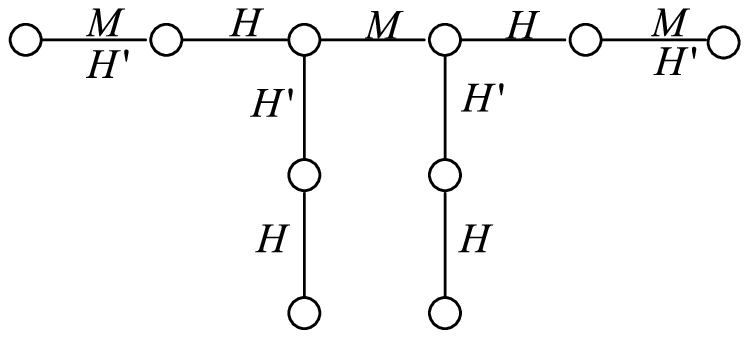}\\
\caption{}\label{fig_alt_path_spanner}
\end{center}
\end{figure}

This lemma implies that each path $P$ from $MP_o^M(M,H)$ together
with the two paths from $Y(M,H,H')$ starting from the end-edges of
$P$ form a spanner (figure \ref{fig_alt_path_spanner}).

Since there are $\beta(G) - \alpha(G)$ paths in $MP_o^M(M,H)$
(property \ref{maxmatchingproperty}), we get:
\begin{corollary}\label{subgraph_F}
There is a subgraph $F$ of the graph $G$ that is an $S$-forest
containing $\beta(G) - \alpha(G)$ spanners as its connected
components.
\end{corollary}

 Now, let $F$ be an $S$-forest arbitrarily chosen among the ones described in the corollary \ref{subgraph_F}.
 Due to property \ref{S-forest_lambda=2alpha=8k}, $\alpha(F) = \lambda(F) - \alpha(F) = 4(\beta(G) -
 \alpha(G))$, therefore due to equality (\ref{fivefourthequality}), $\alpha(F) = \lambda(F) - \alpha(F) = \alpha(G) = \lambda(G) -
 \alpha(G)$. This means that $M_2(F) \subseteq M_2(G)$.

\bigskip

Let $(H, H')$ be an arbitrarily chosen pair from $M_2(F) \subseteq
M_2(G)$. Note that the choice of $(H,H')$ differs from the one above
(we keep this notation as the reader may have already got used with
a pair from $M_2(G)$ denoted by $(H,H')$).

\vspace*{0.4cm}

\begin{lemma} \label{mandatory3-vertex}
If $(u,v)\in E(G)\backslash E(F)$, and $u \notin V(F)$ or $u$ is a
$1$-vertex of $F$, then $v$ is a $3$-vertex of $F$.
\end{lemma}
\begin{proof}
Due to property \ref{missedbyone}, without loss of generality, we
may assume that $u$ is missed by $H'$.

Clearly, $v \in V(F)$ as otherwise $v$ would also be missed by $H'$
and we could ``enlarge'' $H'$ by ``adding'' $(u, v)$ to it, which
contradicts $(H, H') \in M_2(G)$.

\begin{figure}[h]
\begin{center}
\includegraphics{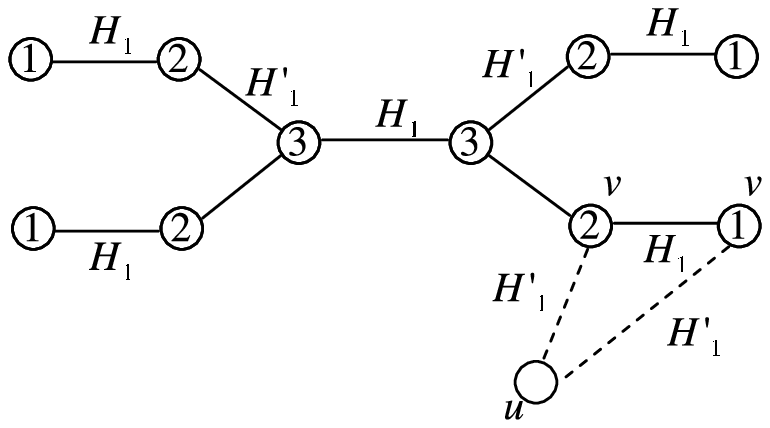}\\
\caption{$S_0$}\label{fig_claim_mandatory3vertex}
\end{center}
\end{figure}

Now, let us show that $v$ is neither a $1$-vertex nor a $2$-vertex.
Suppose for contradiction that it is, and let $S_0$ be the spanner
(connected component) of $F$ containing $v$. Define matchings $H_1,
H_1'$ as follows (figure \ref{fig_claim_mandatory3vertex}):
$$H_1 \doteq (H \backslash E(S_0)) \cup M_0,$$
where $M_0$ is the perfect matching of $S_0$;

$$H_1' \doteq (H' \backslash E(S_0)) \cup J_0,$$
where $J_0$ is a matching of cardinality three satisfying $J_0
\subseteq L(S_0) \cup \{(u,v)\}$ (it always exists). Clearly, $H_1
\cap H_1' = \emptyset$, and, since $|H \cap E(S_0)| = |H' \cap
E(S_0)| = 4$, $|M_0| = 5$ and $|J_0| = 3$, we have $|H_1| + |H_1'| =
(|H| - 4 + 5) + (|H'| - 4 + 3) = \lambda(G)$ and $|H_1|
> |H|$. This contradicts $(H,H') \in M_2(G)$, concluding the proof
of the lemma.
\end{proof}

\begin{lemma}\label{cannotbe3-vertex}
If $(u,v) \in E(G) \backslash E(F)$, then $u \in V(F)$ and if $u$ is
a $1$-vertex of $F$ then $v$ is its base.
\end{lemma}
\begin{proof}
Assume the contrary. Let $(u,v) \in E(G) \backslash E(F)$, where $u$
either belongs to $V(G) \backslash V(F)$, or is a $1$-vertex whose
base is not $v$. As $(u,v)$ satisfies the conditions of the lemma
\ref{mandatory3-vertex}, implies that $v$ is a $3$-vertex of $F$.
Let $W_0$ be the side of the spanner $S_0$ (connected component of
$F$) containing $v$. It is easy to notice that $u$ does not belong
to $V(W_0)$ as otherwise it would be a $1$-vertex of $S_0$ whose
base is $v$. Due to property \ref{missedbyone}, without loss of
generality, we may assume that $u$ is missed by $H'$. Define
matchings $H_1$ and $H_1'$ as follows (figure
\ref{fig_claim_cannotbe3-vertex}):
$$H_1 \doteq (H \backslash E(W_0)) \cup \{ (u,v) \} \cup
(U(S_0) \cap E(W_0)),$$
$$H_1' \doteq (H' \backslash E(W_0)) \cup \{ e \},$$
where $e$ is an edge from $L(S_0 \cap E(W_0)$.

\begin{figure}[h]
\begin{center}
\includegraphics{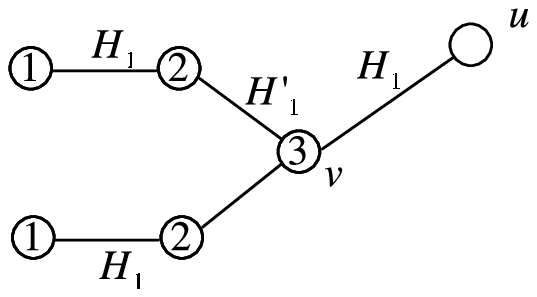}\\
\caption{$W_0$}\label{fig_claim_cannotbe3-vertex}
\end{center}
\end{figure}

Clearly, $H_1$ and $H_1'$ are disjoint matchings. Moreover, since
$|H \cap E(W_0)| = |H' \cap E(W_0)| = 2$, $|H_1| + |H_1'| = (|H| - 2
+ 1 + 2) + (|H'| - 2 + 1) = \lambda(G)$ and $|H_1| > |H|$, which
contradicts $(H,H') \in M_2(G)$ concluding the proof of the lemma.
\end{proof}

\begin{lemma} \label{S-graph}
$G$ is an $S$-graph with spanning $S$-forest $F$.
\end{lemma}
\begin{proof}
Lemma \ref{cannotbe3-vertex} asserts that there is no edge incident
to a vertex from $V(G) \backslash V(F)$, i.e. all vertices from
$V(G) \backslash V(F)$ are isolated. This is a contradiction as we
assume that $G$ has no isolated vertices (see the beginning of
Introduction). Thus, $V(G) \backslash V(F) = \emptyset$ and $F$ is a
spanning $S$-forest of $G$, which means that $G$ is an $S$-graph.
\end{proof}

Due to this lemma, $F$ (an arbitrarily chosen $S$-forest with
$\beta(G) - \alpha(G)$ spanners) is spanning. Obviously, the
converse is also true. So, we may say that $F$ is an arbitrary
spanning $S$-forest of $G$.

\begin{lemma} \label{condition1}
The graph $G$ with its spanning $S$-forest $F$ satisfies the
condition (a) of the theorem.
\end{lemma}
\begin{proof}
Let $u$ be a $1$-vertex of $F$. Lemma \ref{cannotbe3-vertex} asserts
that if $e = (u,v)$ is an edge from $E(G) \backslash E(F)$ then $v$
is the base of $u$, thus $e \in \Delta(G, F)$. This means that the
condition (a) is satisfied.
\end{proof}

\begin{lemma} \label{condition2}
The graph $G$ with its spanning $S$-forest $F$ satisfies the
condition (b) of the theorem.
\end{lemma}
\begin{proof}
Assume that $u$ is a $1$-vertex of $F$ incident to an edge $(u, w)$
from $\Delta(G, F)$ ($w$ is the base of $u$), and $v$ is the
$2$-vertex of $F$ adjacent to $u$ (this $2$-vertex is unique as, due
to lemma \ref{condition1}, $u$ can be incident only to edges from
$U(F)$ and $\Delta(G, F)$). On the opposite assumption $v$ is
incident to an edge $(v, w')$ from $B(G, F)$ (figure
\ref{fig_2_stuff}a).

\begin{figure}[h]
\begin{center}
\includegraphics{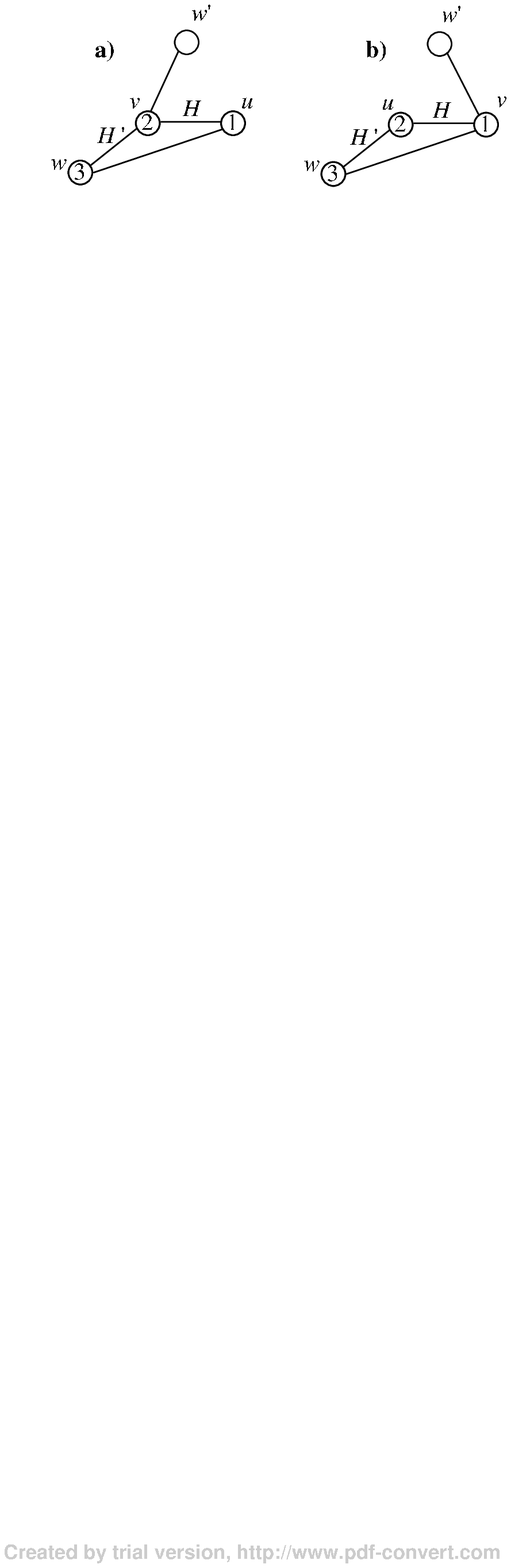}\\
\caption{}\label{fig_2_stuff}
\end{center}
\end{figure}

Let us construct a subgraph $F'$ of $G$ by removing $(v, w)$ from
$F$ and adding $(w, u)$:
$$F' \doteq (F \backslash \{(v, w)\}) \cup \{(w, u)\}.$$
Note that $F'$ is a spanning $S$-forest of $G$, for which $v$ is a
$1$-vertex (figure \ref{fig_2_stuff}b), and $B(G, F') = B(G, F)$.
Thus, $(v, w') \in B(G, F')$. On the other hand, the graph $G$ with
its spanning $S$-forest $F'$ satisfies the condition (a) of the
theorem (lemma \ref{condition1}). Thus, $v$ cannot be incident to an
edge from $B(G, F')$, and we have a contradiction.
\end{proof}

\begin{lemma} \label{condition3}
The graph $G$ with its spanning $S$-forest $F$ satisfies the
condition (c) of the theorem.
\end{lemma}
\begin{proof}
Suppose for contradiction that there exists an $L(F)$-$B(G,F)$
alternating even cycle $C$ of the graph $G$ containing a $2$-$2$
edge $e_0$ and the graph $C_{B(G,F)}$ is bipartite.

The definition of $L(F)$ implies that for each $v \in V(C)$, $1 \leq
d_{C_{L(F)}}(v) \leq 2$. Therefore, due to property
\ref{altpathequaldegrees}, $1 \leq d_{C_{B(G,F)}}(v) \leq 2$, and if
$v$ is a $2$-vertex, then $d_{C_{B(G,F)}}(v) = 1$. Hence, the
connected components of the bipartite graph $C_{B(G,F)}$ are paths
or cycles of even length.

Choose $(H_1,H_1') \in M_2(C_{B(G,F)})$. Due to property
\ref{path_evencycle_HH'}, $H_1 \cup H_1' = E(C_{B(G,F)})$.

Let $F_0$ be the subgraph of $F$ whose connected components are
those sides of the spanners (connected components) $S_1, S_2, ...,
S_{\beta(G) - \alpha(G)}$ of $F$, which do not contain any edge from
$C$. Note that none of the edges of $F_0$ is adjacent to an edge
lying on $C$ since otherwise one of the edges from $L(F) \cap
E(F_0)$ would lie on $C$ as well contradicting the definition of
$F_0$. Let $(H_2, H_2') \in M_2(F_0)$. As the sides of spanners are
paths, again due to property \ref{path_evencycle_HH'}, $H_2 \cup
H_2' = E(F_0)$.

Define the sets $U', L', U''$ as follows:
$$U' \doteq \{ e \in U(F) : e \text{ is adjacent to an edge } e' \in E(C_{B(G,F)}) \},$$
$$L' \doteq \{ e \in L(F) : e \notin C, \text{ $e$ is adjacent to an edge } e'
\in E(C_{B(G,F)}) \},$$
$$U'' \doteq \{ e \in U(F) : \text{$e$ is adjacent to an edge } e' \in L' \}.$$
Define a pair of disjoint matchings $(H_3,H_3')$ as follows:
$$H_3 \doteq \{ e \in U' \cup L' : \text{the edge $e' \in E(C_{B(G,F)})$ adjacent
to $e$ does not belong to } H_1 \},$$
$$H_3' \doteq \{ e \in U' \cup L' : \text{the edge $e' \in E(C_{B(G,F)})$
adjacent to $e$ does not belong to } H_1' \},$$ Note that $H_3 \cup
H_3' = U' \cup L'$.

Define a pair of disjoint matchings $(H_4,H_4')$ as follows:
$$H_4 \doteq \{ e \in U'' : \text{the edge $e' \in L'$ adjacent to $e$
does not belong to } H_3 \},$$
$$H_4' \doteq \{ e \in U'' : \text{the edge $e' \in L'$ adjacent to $e$
does not belong to } H_3' \},$$ Note that $H_4 \cup H_4' = U''$.

Finally, define a pair of disjoint matchings $(H_0,H_0')$ as
follows:
$$H_0 \doteq \bigcup_{i=1}^4 H_i, \ H_0' \doteq \bigcup_{i=1}^4 H_i'.$$
Since $U(F) = (E(F_0) \cap U(F)) \cup U' \cup U''$ and $L(F) =
(E(F_0) \cap L(F)) \cup L' \cup E(C_{L(F)})$, we have $$H_0 \cup
H_0' = U(F) \cup (L(F) \backslash E(C_{L(F)})) \cup E(C_{B(G,
F)}).$$ Hence,
$$|H_0| + |H_0'| = |H_0 \cup H_0'| = |U(F)| + |L(F)| - |E(C_{L(F)})| +
|E(C_{B(G,F)})| = |U(F)| + |L(F)|$$
since $|E(C_{L(F)})| = |E(C_{B(G,F)})|$. \\
From the equality (\ref{fivefourthequality}) we have that
$\lambda(G) = 8(\beta(G) - \alpha(G))$. Therefore, as $|U(F)| +
|L(F)| = 8(\beta(G) - \alpha(G))$, we get:
$$|H_0| + |H_0'| = \lambda(G).$$
As $\lambda(G) = 2 \alpha(G)$ (equality (\ref{fivefourthequality})),
due to property \ref{lambda=2alpha}, there is no maximal
$H_0$-$H_0'$ alternating odd path ($MP_o(H_0, H_0') = \emptyset$).
Let us show that there is one and get a contradiction.

Let $e_1, e_2$ be the edges from $U(F)$ adjacent to $e_0$. Clearly
$e_1, e_2 \in U'$ as $e_0 \in E(C_{B(G, F)})$. The construction of
$H_0$ and $H_0'$ (or rather $H_3$ and $H_3'$) implies that the path
$e_1, e_0, e_2$ is an $H_0$-$H_0'$ alternating odd path. Let $u$ be
the $1$-vertex incident to $e_1$. Lemmas \ref{condition1} and
\ref{condition2} imply that $u$ is not incident to an edge other
than $e_1$. The same can be shown for $e_2$. Therefore, the path
$e_1, e_0, e_2$ is a maximal $H_0$-$H_0'$ alternating path of odd
length (belongs to $MP_o(H_0, H_0')$). This contradiction concludes
the proof of the lemma.
\end{proof}

\bigskip
\bigskip

Lemmas \ref{S-graph}, \ref{condition1}, \ref{condition2} and
\ref{condition3} imply the following statement, which is stronger
than the Necessity of the theorem:

\begin{statement}\label{necessity}
If for a graph $G$ the equality
$\frac{\beta(G)}{\alpha(G)}=\frac{5}{4}$ holds, then $G$ is an
$S$-graph \textbf{any} spanning $S$-forest $F$ of which satisfies
the conditions (a), (b) and (c) of the theorem.
\end{statement}

\subsection{Sufficiency}

The structure of the proof of the Sufficiency is the following: for
an $S$-graph $G$ with a spanning $S$-forest $F$ satisfying the
conditions (a) and (b) of the theorem we show that if
$\frac{\beta(G)}{\alpha(G)} \neq \frac{5}{4}$ then the condition (c)
is not satisfied.

The proof of the Sufficiency is more complicated than the one of the
Necessity. Therefore, we first present the idea of the proof
briefly.

Let $G$ be an $S$-graph with a spanning $S$-forest $F$ satisfying
the condition (a) of the theorem. Let us make also an additional
assumption:

\begin{assumption}\label{assumption_no_delta}
There is a pair $(S, S') \in M_2(G) \backslash M_2(F)$ such that $(S
\cup S') \cap \Delta(G, F) = \emptyset$.
\end{assumption}

Choose an arbitrary pair $(H, H')$ from $M_2(F)$.

Let $T_U, T_L, T_B$ be the sets of edges from $S' \backslash H'$
that belong to $U(F), L(F), B(G,F)$ respectively. Note that $T_U,
T_L, T_B$ are pairwise disjoint, and, due to assumption
\ref{assumption_no_delta}, we have:
$$S' \backslash H' = T_U \cup T_L \cup T_B.$$

Define $Q, Q_U, Q_L, Q_B$ as the sets of all vertices that are
incident to edges from $S' \backslash H'$, $T_U$, $T_L$, $ T_B$,
respectively. Note that $Q_U, Q_L, Q_B$ are pairwise disjoint, $|Q|
= 2|S' \backslash H'|$, and
$$Q = Q_U \cup Q_L \cup Q_B.$$

Using paths from $MP(S, H)$ we construct a set of trails $A''
\subseteq T_o^{L(F)}(L(F), B(G,F))$ such that all edges lying on
trails from $A''$ that belong to $B(G, F)$ are from $S$, and
$V_0(A'') = Q_B$. Note that the trails from $A''$ are connected with
edges from $T_B$ (as shown in figure \ref{fig_LB_alt_cycle}) making
$L(F)$-$B(G,F)$ alternating even cycles, i.e. each edge from $T_B$
lies on one such cycle. Also note that for any such cycle $C$, the
graph $C_{B(G,F)}$ does not contain odd cycles as $E(C_{B(G,F)})
\subseteq S \cup S'$, therefore $C_{B(G,F)}$ is bipartite.

After this, assuming that the condition (b) is also satisfied and
$\frac{\beta(G)}{\alpha(G)} \neq \frac{5}{4}$ (it is shown that
these assumptions together are stronger than assumption
\ref{assumption_no_delta}), we prove that $T_B$ contains a $2$-$2$
edge. Therefore, at least one of the $L(F)$-$B(G,F)$ alternating
even cycles described above contains a $2$-$2$ edge contradicting
the condition (c) of the theorem.

The construction of $A''$ is a step-by-step process. First, from
$MP(S, H)$ we construct a set of paths $A$ for which $Q_U \cup Q_L
\cup Q_B \subseteq V_2(A)$. Then, $A$ is transformed to a set of
trails $A'$ for which $V_0(A') = Q_L \cup Q_B$. And finally, $A'$ is
transformed to $A''$ mentioned above.

\bigskip

Now, let us start the proof.

As mentioned above, assume that $G$ is an $S$-graph with a spanning
$S$-forest $F$ satisfying the condition (a) of the theorem, and the
assumption \ref{assumption_no_delta} holds. Choose $(S, S')$ and
$(H, H')$ as described above.

In order to characterize the set $Q$ define the sets $R_1, R_2, R_3,
R$ as follows:
$$R_1 \doteq \{v \in V(G) : v \text{ is incident to an edge from } H'
\backslash S'\},$$
$$R_2 \doteq \{v \in R_1 : v \text{ is a 1-vertex of } F\},$$
$$R_3 \doteq \{v \in V(G) : v \text{ is a 1-vertex of $F$ incident
to an edge from } H \backslash S\},$$
$$R \doteq (R_1 \backslash R_2) \cup R_3.$$

We claim that $R$ is a set of vertices ``potentially" incident to
edges from $S' \backslash H'$. Formally:
\begin{lemma}\label{QinR}
$Q \subseteq R$.
\end{lemma}
\begin{proof}
Assume that $v \in Q$ and $e$ is the edge from $S' \backslash H'$
incident to $v$. If $v$ is not a $1$-vertex then, due to property
\ref{coveredbyboth}, $v$ is incident to an edge from $H' \backslash
S'$, and therefore, belongs to $R_1 \backslash R_2$. On the other
hand, if $v$ is a $1$-vertex then, due to condition (a) and
assumption \ref{assumption_no_delta}, $e \in U(F)$. Moreover, $e \in
H \backslash S$ as $e$ belongs to $S'$ and does not belong to $H'$.
Thus, $v \in R_3$.
\end{proof}

Further, we show that in fact $Q = R$. We introduce $R$ as its
definition is much easier to work with.

Consider the paths from $MP(S,H)$ (they are the main working tools
throughout the whole proof).

\bigskip

From condition (a) of the theorem and assumption
\ref{assumption_no_delta} we get the following corollaries:
\begin{corollary} \label{one_is_not_incident_to_two}
Any $1$-vertex of $F$ is incident to at most one edge from $S
\triangle H$.
\end{corollary}

\begin{corollary} \label{U_on_paths}\
\renewcommand{\labelenumi}{(\arabic{enumi})}
\begin{enumerate}
\item \label{one_is_end_vertex} If a $1$-vertex (an edge from $U(F)$)
lies on a path from $MP(S,H)$, then it is an end-vertex (end-edge)
of the latter.

\item \label{no_U_lies_on_cycles} No $1$-vertex (edge from $U(F)$) lies on a cycle
from $C_e(S,H)$.
\end{enumerate}
\end{corollary}

\begin{lemma} \label{S_on_U}
If $e \in S$ is an end-edge of a path $P \in MP(S,H)$ and $v$ is an
end-vertex of $P$ incident to $e$, then $e \in U(F)$ and $v$ is a
$1$-vertex of $F$.
\end{lemma}
\begin{proof}
Let $e_1 \in S$ be an end-edge of a path $P : e_1, e_2, ..., e_n \in
MP(S,H)$, and $v$ be an end-vertex of $P$ incident to $e_1$. If $v$
is not a $1$-vertex then, due to property \ref{coveredbyboth}, $v$
is incident to an edge $e \in H$. Due to the definition of
alternating path, $e \neq e_1$, therefore the path $e, e_1, e_2,
..., e_n$ belongs to $P(S, H)$, which contradicts the maximality of
$P$ (see the definition \ref{max_alt_path_definition} of a maximal
path).
\end{proof}

\bigskip

As a corollary from lemma \ref{S_on_U} we get:
\begin{corollary}\label{odd_S_paths_length}
Every path from $MP_o^S(S,H)$ is of length at least five.
\end{corollary}

\begin{lemma}\label{cardinality_ScapU}
$|(S \backslash H) \cap U(F)| = 2|MP_o^S(S,H)| + |MP_e(S,H)|$.
\end{lemma}
\begin{proof}
Due to property \ref{edge_lies_on_alt_component}, every edge from $S
\backslash H$ lies either on a path from $MP(S,H)$ or a cycle from
$C_e(S,H)$. Moreover, corollary \ref{U_on_paths} implies that every
edge from $(S \backslash H) \cap U(F)$ is an end-edge of a path from
$MP(S,H)$, hence $|(S \backslash H) \cap U(F)| \leq 2|MP_o^S(S, H)|
+ |MP_e(S, H)|$. Corollary \ref{odd_S_paths_length} implies that
each path from $MP_o^S(S, H)$ has two different end-edges. Thus, due
to lemma \ref{S_on_U}, $|(S \backslash H) \cap U(F)| \geq
2|MP_o^S(S, H)| + |MP_e(S, H)|$, which completes the proof of the
lemma.
\end{proof}

\bigskip

The following two lemmas provide us with three inequalities, the
boundary cases (equalities) of which are related to a number of
useful properties. Those inequalities together imply that the
equalities are indeed the case.

\begin{lemma}\label{balance_upper_bound}
$2|H' \backslash S'| - |R| \leq 2|H' \backslash S'| - |Q| \leq
2(|MP_o^S(S,H)| - MP_o^H(S,H)|)$, and the equality cases in the
first and second inequalities hold if and only if $Q = R$ and
$\lambda(F) = \lambda(G)$, respectively.
\end{lemma}
\begin{proof}
By the definition of $\lambda$, $\lambda(F) = |H| + |H'| \leq |S| +
|S'| = \lambda(G)$. Due to property \ref{cardinalitydiff}, this is
equivalent to
$$|H'| - |S'| \leq |S| - |H| = |MP_o^S(S,H)| - MP_o^H(S,H)|,$$
or
$$|H' \backslash S'| - |S' \backslash H'| \leq |MP_o^S(S,H)| - MP_o^H(S,H)|.$$
Taking into account that $2|S' \backslash H'| = |Q|$ we get:
$$2|H' \backslash S'| - |Q| \leq 2(|MP_o^S(S,H)| - MP_o^H(S,H)|),$$
and the equality holds if and only if $\lambda(F) = \lambda(G)$.

Furthermore, lemma \ref{QinR} implies that $|Q| \leq |R|$, and we
get
$$2|H' \backslash S'| - |R| \leq 2|H'
\backslash S'| - |Q|,$$ and the equality holds if and only if $|Q| =
|R|$, which means that $Q = R$, due to lemma \ref{QinR}.
\end{proof}

\begin{lemma}\label{balance_lower_bound}
$2|H' \backslash S'| - |R| \geq |(H' \backslash (S \cup S')) \cap
U(F)| + 2(|MP_o^S(S,H)| - MP_o^H(S,H)|)$, and the equality case
holds if and only if for any path from $MP(S,H)$, if its end-edge $e
\in H \backslash S$ then $e \in U(F)$.
\end{lemma}
\begin{proof}
First note that
$$|R_1| = 2|H' \backslash S'|.$$

Clearly, any vertex from $R_2$ is incident to one edge from $(H'
\backslash S') \cap U(F)$ and vice versa. Furthermore, any edge from
$(H' \backslash S') \cap U(F)$ either belongs to $(H' \backslash (S
\cup S')) \cap U(F)$, or belongs to $(S \backslash H) \cap U(F)$.
Thus, due to lemma \ref{cardinality_ScapU}, we get:
$$|R_2| = |(H' \backslash (S \cup S')) \cap U(F)| + |(S
\backslash H) \cap U(F)| =$$$$= |(H' \backslash (S \cup S')) \cap
U(F)| + 2|MP_o^S(S,H)| + |MP_e(S,H)|.$$

Moreover, property \ref{edge_lies_on_alt_component} together with
corollary \ref{U_on_paths} implies that every vertex $v \in R_3$ is
an end-vertex of a path from $MP(S,H)$, and the end-edge of that
path incident to $v$ is from $H \backslash S$. Thus,
\setcounter{equation}{0}
$$|R_3| \leq 2|MP_o^H(S,H)| + |MP_e(S,H)|,$$
and the equality holds if and only if the vice versa is also true,
i.e. any end-edge $e \in H \backslash S$ of a path from $MP(S,H)$ is
from $U(F)$.

Hence we have
$$|R| = |R_1| - |R_2| + |R_3| \leq 2|H' \backslash S'| - (|(H' \backslash (S \cup S')) \cap U(F)|
+ 2|MP_o^S(S,H)| + |MP_e(S,H)|) +$$$$ + 2|MP_o^H(S,H)| +
|MP_e(S,H)|,$$ or
$$2|H' \backslash S'| - |R| \geq |(H' \backslash (S \cup S')) \cap U(F)|
+ 2(|MP_o^S(S,H)| - MP_o^H(S,H)|).$$
\end{proof}

\bigskip

Lemmas \ref{balance_upper_bound} and \ref{balance_lower_bound}
together imply the following:
\begin{corollary}\label{balance_corollaries}\
\renewcommand{\labelenumi}{(\arabic{enumi})}
\begin{enumerate}
\item \label{balance_equality}
$2|H' \backslash S'| - |R| = 2(|MP_o^S(S,H)| - MP_o^H(S,H)|);$

\item \label{Q=R}
$Q = R;$

\item \label{lambda_remains_the_same}
$\lambda(F) = |H| + |H'| = |S| + |S'| = \lambda(G);$

\item \label{H_on_U}
If $e \in H \backslash S$ is an end-edge of a path from $MP(S,H)$
and $v$ is an end-vertex incident to $e$ then $e \in U(F)$ and $v$
is a $1$-vertex of $F$;

\item \label{clear_H'_not_in_U}
$(H' \backslash (S \cup S')) \cap U(F) = \emptyset.$
\end{enumerate}
\end{corollary}

\bigskip

Statement (\ref{clear_H'_not_in_U}) of corollary
\ref{balance_corollaries} and property \ref{S-forest_HH'_is_LU}
imply:
\begin{corollary}\label{clear_H'_corollaries}\
\renewcommand{\labelenumi}{(\arabic{enumi})}
\begin{enumerate}
\item \label{clear_H' in_L}
$H' \backslash (S \cup S') \subseteq L(F);$

\item \label{clear_H'_inc_vertex_in_R}
Vertices incident to edges from $H' \backslash (S \cup S')$ belong
to $R_1 \backslash R_2 \subseteq R = Q$.
\end{enumerate}
\end{corollary}

\bigskip

Statement (\ref{H_on_U}) of corollary \ref{balance_corollaries} and
lemma \ref{S_on_U} imply the following:
\begin{corollary}\label{paths_corollaries}\
\renewcommand{\labelenumi}{(\arabic{enumi})}
\begin{enumerate}
\item \label{length_at_least_3}
The length of any path from $MP_o^H(S,H)$ is at least three;

\item \label{even_length_at_least_4}
The length of any path from $MP_e(S,H)$ is at least four;

\item \label{every_path_is 1-1}
Every path of $MP(S,H)$ connects $1$-vertices (end-vertices are
$1$-vertices).
\end{enumerate}
\end{corollary}

\bigskip

Now we are able to give a characterization of the set $T_U$.
\begin{lemma} \label{H_in_U_is_TU}
$T_U = (H \backslash S) \cap U(F)$.
\end{lemma}
\begin{proof}
Let $e \in (H \backslash S) \cap U(F)$ and $v$ be the $1$-vertex
incident to $e$. The definition of $R_3$ implies that $v \in R_3
\subseteq R$, and due to statement (\ref{Q=R}) of the corollary
\ref{balance_corollaries}, $v \in Q$. As $v$ is a 1-vertex, $v \in
Q_U$, thus, $e \in T_U$. Hence, $(H \backslash S) \cap U(F)
\subseteq T_U$.

On the other hand, by definition $T_U = (S' \backslash H') \cap
U(F)$. For any edge $e \in (S' \backslash H') \cap U(F)$ we have
that $e \notin S$ as $e \in S'$, and $e \in H$ as $e \in U(F)$ and
$e \notin H'$ (property \ref{S-forest_HH'_is_LU}). Thus, $e \in (H
\backslash S) \cap U(F)$, i.e. $T_U \subseteq (H \backslash S) \cap
U(F)$ and the proof of the lemma is complete.
\end{proof}

\bigskip

The following lemma describes the placement of edges from $H'$ lying
on $S$-$H$ alternating trails (maximal paths or simple even cycles):
\begin{lemma} \label{H'on_paths}\
\renewcommand{\labelenumi}{(\arabic{enumi})}
\begin{enumerate}
\item \label{end_and_beforeend} If $e \in H'$ lies on a path $P \in
MP(S,H)$ then $e \in E_2(P)$;
\item \label{no_H'_on_cycles} No edge from $H'$ lies on a cycle from
$C_e(S,H)$.
\end{enumerate}
\end{lemma}
\begin{proof}
Let $e \in H'$ and note that if $e$ lies on a path from $MP(S,H)$ or
on a cycle from $C_e(S,H)$ then $e \in S$.

(\ref{end_and_beforeend}). Suppose that $e \notin E_1(P)$. Then, $e
\in L(F)$ as otherwise $e \in U(F)$ (property
\ref{S-forest_HH'_is_LU}) and, due to statement
(\ref{one_is_end_vertex}) of corollary \ref{U_on_paths}, $e \in
E_1(P)$. Therefore, due to maximality of $P$, $e$ is adjacent to two
edges from $H \backslash S$ lying on $P$, one of which (denote it by
$h$) belongs to $U(F)$. Due to statement (\ref{one_is_end_vertex})
of corollary \ref{U_on_paths}, $h \in E_1(P)$, which means that $e
\in E_2(P) \backslash E_1(P)$.

(\ref{no_H'_on_cycles}). Suppose for contradiction that $e$ lies on
a cycle $C \in C_e(S,H)$. Here again, $e \in L(F)$ as otherwise $e
\in U(F)$ (property \ref{S-forest_HH'_is_LU}) contradicting
statement (\ref{no_U_lies_on_cycles}) of corollary \ref{U_on_paths}.
Therefore, there are two edges from $H \backslash S$ adjacent to $e$
lying on $C$, one of which belongs to $U(F)$, which contradicts
statement (\ref{no_U_lies_on_cycles}) of corollary \ref{U_on_paths}.
\end{proof}

\bigskip

Now let us construct the set of paths $A$ mentioned above:
$$A \doteq MP(S,H) \cup (H' \backslash (S \cup S'))$$
(the edges from $H' \backslash (S \cup S')$ are considered as paths
of length one).

The following lemma provides the above-mentioned property of $A$,
for which $A$ is actually constructed.
\begin{lemma} \label{R_on_paths}
$Q \subseteq V_2(A)$.
\end{lemma}
\begin{proof}
Due to statement \ref{Q=R} of corollary \ref{balance_corollaries},
$Q = R$. So, assume that $v \in R_1 \backslash R_2$ and $e$ is the
edge from $H' \backslash S'$ incident to $v$. If $e \notin S$, then
$e \in H' \backslash (S \cup S')$ and we are done. On the other
hand, if $e \in S$, then, as $e \notin H$, $e$ lies on a path $P \in
MP(S,H)$ (due to statement (\ref{no_H'_on_cycles}) of lemma
\ref{H'on_paths}, $e$ cannot lie on a cycle from $C_e(S,H)$). Hence,
according to statement (\ref{end_and_beforeend}) of lemma
\ref{H'on_paths}, $e \in E_2(P)$, which means that $v \in V_2(P)$.

Now let $v \in R_3$ be a $1$-vertex incident to the edge $e \in (H
\backslash S) \cap U(F)$. Clearly, $e$ lies on a path from $MP(S,H)$
as it cannot lie on a cycle from $C_e(S,H)$ (see statement
(\ref{no_U_lies_on_cycles}) of corollary \ref{U_on_paths}).
Moreover, $v$ lies on $P$ too, and, due to statement
(\ref{one_is_end_vertex}) of corollary \ref{U_on_paths}, is an
end-vertex of $P$, concluding the proof of the lemma.
\end{proof}

\bigskip

Now we intend to transform $A$ to a set of trails $A' \subseteq
T_o^{L(F)}(L(F), B(G,F))$, such that $V_0(A')=Q_L \cup Q_B$ and any
edge from $B(G,F)$ lying on a trail from $A'$ belongs to $S$.

For each path $P$ of $MP(S,H)$ we transform only the edges of sets
$E_2^b(P)$ and $E_2^e(P)$ (first two and last two edges of the
path). Note that corollary \ref{odd_S_paths_length} and statements
(\ref{length_at_least_3}) and (\ref{even_length_at_least_4}) of
corollary \ref{paths_corollaries} imply that $E_2^b(P)$ and
$E_2^e(P)$ are sets of cardinality $2$ which do not coincide though
may intersect.

For each path $P \in MP(S,H)$, and for $X=E_2^b(P)$ or $X=E_2^e(P)$,
where $X=\{ e, e' \}$, $e \in E_1(P)$ and $e' \in E_2(P) \backslash
E_1(P)$, do the following:

\begin{case} \label{case_einH_e'inB}
$e \in H, e' \in B(G,F)$ (figure \ref{fig_trans1_1}a).

Due to statement (\ref{H_on_U}) of corollary
\ref{balance_corollaries}, $e \in U(F)$. Let $e''$ be the edge from
$L(F)$ adjacent to $e$ and $e'$ (figure \ref{fig_trans1_1}a).

\begin{figure}[h]
\begin{center}
\includegraphics{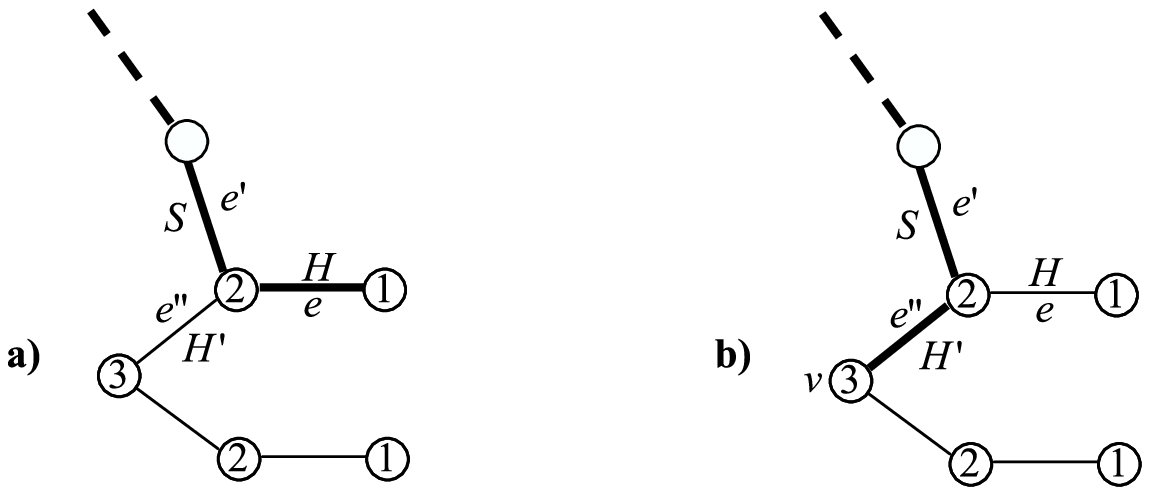}\\
\caption{}\label{fig_trans1_1}
\end{center}
\end{figure}

Remove $e$ from $P$ and add (concatenate) $e''$ instead (figure
\ref{fig_trans1_1}b).
\end{case}

\begin{case} \label{case_einH_e'notinB}
$e \in H, e' \notin B(G,F)$ (hence $e' \in L(F)$) (figure
\ref{fig_trans1_2}a).

\begin{figure}[h]
\begin{center}
\includegraphics{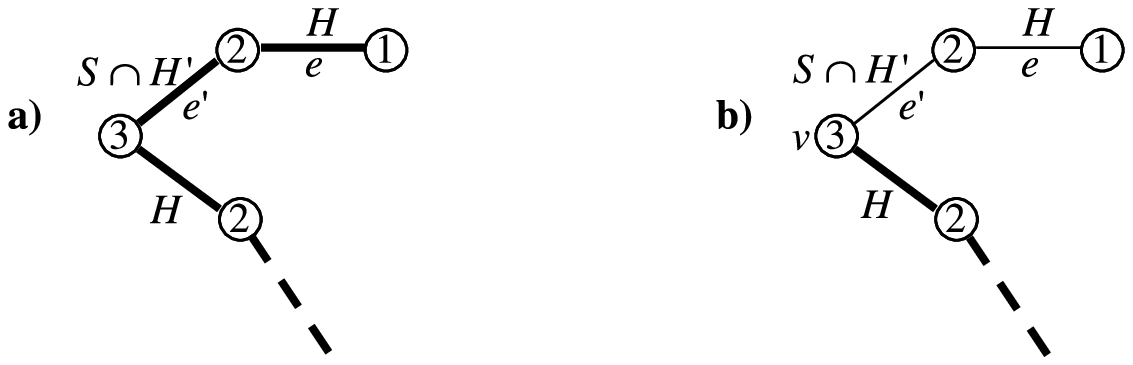}\\
\caption{}\label{fig_trans1_2}
\end{center}
\end{figure}

Here again due to statement (\ref{H_on_U}) of corollary
\ref{balance_corollaries}, $e \in U(F)$. Remove $e$ and $e'$ from
$P$ (figure \ref{fig_trans1_2}b).
\end{case}

\begin{case} \label{case_einS}
$e \in S$.

As $e \in U(F)$ (lemma \ref{S_on_U}) and $e' \in H$ ($e \in S$), $e'
\in L(F)$ (figure \ref{fig_trans1_3}a).

\begin{figure}[h]
\begin{center}
\includegraphics{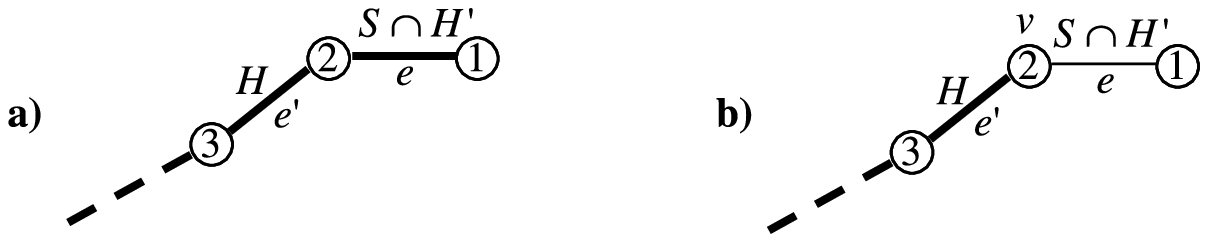}\\
\caption{}\label{fig_trans1_3}
\end{center}
\end{figure}

Remove $e$ from $P$ (figure \ref{fig_trans1_3}b).
\end{case}

Note that the transformation described above is defined correctly
because of the following: for a path $P \in MP(S,H)$, $E_2^b(P)$ and
$E_2^e(P)$ may have non-empty intersection only if $P \in
MP_o^H(S,H)$ and the length of $P$ is three. In this case both
$E_2^b(P)$ and $E_2^e(P)$ are handled by the case
\ref{case_einH_e'inB}, and the edge $e' \in E_2^b(P) \cap E_2^e(P)$
is treated in the same way.

Define sets of edges $Z_{Q_U}$ and $\overline{Z_{Q_U}}$ as follows:
$$Z_{Q_U} \doteq \{ e \in H' \backslash (S \cup S') \text{ / $e$ is incident to a vertex from $Q_U$} \},$$
$$\overline{Z_{Q_U}} \doteq (H' \backslash (S \cup S')) \backslash Z_{Q_U},$$
and let $D$ be the set of trails obtained from the paths of
$MP(S,H)$ by the transformation described above (we do not say that
$D$ is a set of paths as it is not hard to construct an example of
$D$ containing a trail that is not a path using the fact that vertex
$v$ in case \ref{case_einH_e'inB} may also lie on $P$).

Let $A'$ be the following:
$$A' \doteq D \cup \overline{Z_{Q_U}}.$$

\bigskip

Some properties of $A'$ are given below.

\begin{lemma} \label{S_cap_H'_is_empty}
No edge from $S \cap H'$ lies on a trail from $A'$.
\end{lemma}
\begin{proof}
First note that no edge from $\overline{Z_{Q_U}}$ belongs to $S$. So
we need to examine only $D$. Statement (\ref{end_and_beforeend}) of
lemma \ref{H'on_paths} and the transformation described above imply
that the only edges lying on trails from $D$ that belong to $H'$ are
edges denoted by $e''$ in the case \ref{case_einH_e'inB} (figure
\ref{fig_trans1_1}). But that edges do not belong to $S$, and the
proof of the lemma is complete.
\end{proof}

\begin{lemma} \label{S_and_B_are_the_same}
If an edge $e$ lies on a trail from $A'$  and $e \in B(G,F) \cup S$
then $e \in B(G,F) \cap S$ \footnote{In other words, the set of
edges from $B(G,F)$ lying on trails from $A'$ and the set of edges
from $S$ lying on trails from $A'$ coincide.}.
\end{lemma}
\begin{proof}
Clearly, all edges from $B(G,F)$ lying on trails from $A'$, lie on
trails from $D$ as $\overline{Z_{Q_U}} \subseteq L(F)$ (statement
(\ref{clear_H' in_L}) of corollary \ref{clear_H'_corollaries}).
First note that all edges from $B(G,F)$ lying on paths from
$MP(S,H)$ belong to $S$ as edges from $H$ belong to $U(F) \cup L(F)$
(property \ref{S-forest_HH'_is_LU}). During the transformation of
$MP(S,H)$ to $D$, the only edges we add are edges $e''$ in case
\ref{case_einH_e'inB} (figure \ref{fig_trans1_1}), which are from
$L(F)$. Thus, still all edges from $B(G,F)$ lying on $D$ belong to
$S$, and the proof of the first part of the lemma is complete.

Let edge $e \in S$ lies on a trail $T \in A'$. Obviously, $T \in D$.
Due to assumption \ref{assumption_no_delta} $S \subseteq U(F) \cup
L(F) \cup B(G,F)$. As we do not add any edge from $H \cap S$ during
the transformation of $MP(S,H)$ to $D$, $e$ does not belong to $H$.
Due to lemma \ref{S_cap_H'_is_empty}, $e$ does not belong to $H'$
either. Therefore, as $H \cup H'$ = $U(F) \cup L(F)$ (property
\ref{S-forest_HH'_is_LU}), $e \in B(G,F)$.
\end{proof}

\begin{lemma} \label{LB_alternating_paths}
$A' \subseteq T_o^{L(F)}(L(F), B(G,F))$.
\end{lemma}
\begin{proof}
Due to statement (\ref{clear_H' in_L}) of corollary
\ref{clear_H'_corollaries}, $\overline{Z_{Q_U}} \subseteq
T_o^{L(F)}(L(F), B(G,F))$, so we need to show the same for $D$ only.
Corollary \ref{U_on_paths} and the construction of $D$ imply that no
edge from $U(F)$ lies on a trail from $D$, therefore every edge from
$H$ lying on a trail from $D$ belongs to $L(F)$ (property
\ref{S-forest_HH'_is_LU}). On the other hand, due to lemma
\ref{S_and_B_are_the_same}, all edges from $S$ lying on trails from
$D$ are from $B(G,F)$. The only edges that do not belong to $S \cup
B(G,F)$ are edges $e''$ in the case \ref{case_einH_e'inB}, which are
from $L(F)$ and are adjacent to edges from $B(G,F)$ (figure
\ref{fig_trans1_1}b). All these together imply $D \subseteq P(L(F),
B(G,F))$. Moreover, the construction of $D$ implies that the edges
from $E_1(D)$ belong to $L(F)$. Therefore, $D \subseteq
T_o^{L(F)}(L(F), B(G,F))$.
\end{proof}

\begin{lemma} \label{endvertices_of_A'_in_Q_LcupQ_B}
$V_0(A') \subseteq Q_L \cup Q_B$.
\end{lemma}
\begin{proof}
Let $v$ be an end-vertex of a trail $P'$ from $A'$. If $P' \in
\overline{Z_{Q_U}}$ then, due to statement
\ref{clear_H'_inc_vertex_in_R} of corollary
\ref{clear_H'_corollaries} and the definition of
$\overline{Z_{Q_U}}$, $v \in Q \backslash Q_U = Q_L \cup Q_B$.
Therefore, assume that $P' \in D$, and let $P$ be the path from
$MP(S,H)$ corresponding to $P'$. Without loss of generality, we may
assume that $v$ is the starting vertex of $P'$. Let $e, e', e''$ be
as it is shown in the corresponding figure describing each case of
the transformation of $P$ to $P'$.

Assume that the transformation of $E_2^b(P)$ is handled by cases
\ref{case_einH_e'inB} or \ref{case_einH_e'notinB} (figures
\ref{fig_trans1_1}b and figure \ref{fig_trans1_2}b). Lemma
\ref{H_in_U_is_TU} implies that $e \in T_U \subseteq S' \backslash
H'$. Thus, $e'' \in H' \backslash S'$ and $e' \in H' \backslash S'$
for cases \ref{case_einH_e'inB} and \ref{case_einH_e'notinB},
respectively. Therefore, as $v$ is a $3$-vertex (not a $1$-vertex),
$v \in R_1 \backslash R_2 \subseteq R = Q$ (statement (\ref{Q=R}) of
corollary \ref{balance_corollaries}). Moreover, $v \notin Q_U$ as
$Q_U$ contains only $1$-vertices and $2$-vertices. Hence, $v \in Q
\backslash Q_U = Q_L \cup Q_B$.

Now assume that the transformation of $E_2^b(P)$ is handled by case
\ref{case_einS} (figure \ref{fig_trans1_3}b). As $e \in S, e' \in
H$, we have $e \in H' \backslash S'$. Since $v$ is a $2$-vertex (not
a $1$-vertex), $v \in R_1 \backslash R_2 \subseteq R = Q$ (statement
(\ref{Q=R}) of corollary \ref{balance_corollaries}). On the other
hand, due to lemma \ref{H_in_U_is_TU}, $e \notin T_U$ as $e \in S$.
Therefore, $v \notin Q_U$. Thus, $v \in Q \backslash Q_U = Q_L \cup
Q_B$.
\end{proof}

\begin{lemma} \label{Q_LcupQ_B_in_endvertices_of_A'}
$Q_L \cup Q_B \subseteq V_0(A')$.
\end{lemma}
\begin{proof}
Let $v \in Q_L \cup Q_B = Q \backslash Q_U$. The condition (a) of
the theorem implies that $v$ is not a $1$-vertex. Therefore, as $Q =
R$ (statement (\ref{Q=R}) of the corollary
\ref{balance_corollaries}), $v \in R_1 \backslash R_2$, hence there
is an edge $e_0 \in H' \backslash S'$ such that $v$ and $e_0$ are
incident. The following cases are possible:

\renewcommand{\labelenumi}{(\arabic{enumi})}
\begin{enumerate}
\item \textbf{$e_0 \in S$}. Then $e \in S \backslash H$ and $e_0 \in E_2(P)$, where $P$ is a path
from $MP(S,H)$ (lemma \ref{H'on_paths}). Without loss of generality
we may assume that $e_0 \in E_2^b(P)$. Let $P'$ be the trail from
$D$ corresponding to $P$. Consider the following two subcases:
\begin{enumerate}
\item \textbf{$e_0 \in U(F)$}. Then $e_0$ is the starting edge of $P$
(statement (\ref{one_is_end_vertex}) of corollary \ref{U_on_paths}).
Clearly, the transformation of $E_2^b(P)$ is handled by case
\ref{case_einS} (see figure \ref{fig_trans1_2}b, $e_0$ corresponds
to $e$ in the figure). Hence $v$ is the starting vertex of $P'$,
since $v$ is not the $1$-vertex incident to $e_0$.

\item \textbf{$e_0 \notin U(F)$}. Hence $e_0 \in L(F)$ as $e_0 \in H'$ (property \ref{S-forest_HH'_is_LU}).
Due to statement (\ref{one_is_end_vertex}) of corollary
\ref{U_on_paths}, $e_0 \in E_2^b(P) \backslash E_1^b(P)$. Thus, the
transformation of $E_2^b(P)$ is handled by case
\ref{case_einH_e'notinB}. Let $e, e'$ be as in figure
\ref{fig_trans1_2}b, and note that $e_0$ corresponds to $e'$. Lemma
\ref{H_in_U_is_TU} implies that $e \in T_U$, hence the $2$-vertex
incident to $e'=e_0$ belongs to $Q_U$. Since $v \notin Q_U$, $v$ is
the $3$-vertex incident to $e'=e_0$, and therefore, is the starting
vertex of $P'$.
\end{enumerate}

\item \textbf{$e_0 \notin S$}. By the definitions of $Z_{Q_U}$ and $\overline{Z_{Q_U}}$, $e_0 \in Z_{Q_U} \cup
\overline{Z_{Q_U}}$. If $e_0 \in \overline{Z_{Q_U}}$, then we are
done. Therefore, assume that $e_0 \in Z_{Q_U}$. The definition of
$Z_{Q_U}$ implies that there is an edge $e_1 \in T_U$ adjacent to
$e_0$. Due to lemma \ref{H_in_U_is_TU}, $e_1 \in (H \backslash S)
\cap U(F)$, and therefore, $e_1$ is an end-edge of some path $P \in
MP(S,H)$ (property \ref{edge_lies_on_alt_component} and corollary
\ref{U_on_paths}). Without loss of generality, we may assume that
$e_1 \in E_1^b(P)$. Note that $E_2^b(P) \backslash E_1^b(P) \neq
\emptyset$ as any path from $MP(S,H)$ has length at least three
(corollary \ref{odd_S_paths_length} and statements
(\ref{length_at_least_3}) and (\ref{even_length_at_least_4}) of
corollary \ref{paths_corollaries}). Thus, let $e_2 \in E_2^b(P)
\backslash E_1^b(P)$. As $e_1 \in H$, $e_2 \in S$. Let $P'$ be the
trail from $D$ corresponding to $P$. Since $e_0 \in L(F)$ (see
statement (\ref{clear_H' in_L}) of corollary
\ref{clear_H'_corollaries} and the definition of $Z_{Q_U}$) and $e_2
\neq e_0$, we have $e_2 \notin L(F)$ and, due to assumption
\ref{assumption_no_delta}, $e_2 \in B(G,F)$. Hence, the
transformation of $E_2^b(P)$ is handled by case
\ref{case_einH_e'inB}, and the edges $e'', e, e'$ in figure
\ref{fig_trans1_1}b correspond to $e_0, e_1, e_2$, respectively. As
$e=e_1 \in T_U$, the $2$-vertex incident to $e''=e_0$ belongs to
$Q_U$. Therefore, $v$ is the $3$-vertex incident to $e''=e_0$. Thus,
$v$ is the starting vertex of $P'$.
\end{enumerate}
\end{proof}

\bigskip

From lammas \ref{endvertices_of_A'_in_Q_LcupQ_B} and
\ref{Q_LcupQ_B_in_endvertices_of_A'} we get the following corollary:
\begin{corollary}\label{V_0=Q_LcupQ_B}
$V_0(A') = Q_L \cup Q_B$.
\end{corollary}

\bigskip

Let us prove one auxiliary lemma, which is used at the end of the
proof of Sufficiency.
\begin{lemma}\label{2-3}
The difference of the numbers of $2$-vertices and $3$-vertices in
$Q_B$ is $$2(|MP_o^S(S,H)| - |MP_o^H(S,H)|).$$
\end{lemma}
\begin{proof}
Let us denote the difference of the numbers of $2$-vertices and
$3$-vertices in a set of vertices $N$ by $D_{23}(N)$.

As $D_{23}(Q_L) = 0$ (any edge from $L(F)$ is incident to one
$2$-vertex and one $3$-vertex), $D_{23}(Q_B) = D_{23}(Q_L \cup
Q_B)$, and due to corollary \ref{V_0=Q_LcupQ_B}, $D_{23}(Q_B) =
D_{23}(V_0(A'))$.

Note that for any trail $P' \in D$ and its corresponding path $P \in
MP(S,H)$
\begin{itemize}
\item $P'$ starts with a $3$-vertex if and only if $P$ starts
with an edge from $H$ (cases \ref{case_einH_e'inB} and
\ref{case_einH_e'notinB}),
\item $P'$ starts with a $2$-vertex if and only if $P$ starts
with an edge from $S$ (case \ref{case_einS}).
\end{itemize}
Thus, there are $2|MP_o^H(S,H)| + |MP_e(S,H)|$ $3$-vertices and
$2|MP_o^S(S,H)| + |MP_e(S,H)|$ $2$-vertices in $V_0(D)$. As
$\overline{Z_{Q_U}} \subseteq L(F)$ (statement (\ref{clear_H' in_L})
of corollary \ref{clear_H'_corollaries}),
$D_{23}(\overline{Z_{Q_U}}) = 0$. Therefore,
$$D_{23}(Q_B) = D_{23}(V_0(A')) = D_{23}(\overline{Z_{Q_U}}) + D_{23}(D) = D_{23}(D) = 2|MP_o^S(S,H)| + |MP_e(S,H)| -$$
$$- (2|MP_o^H(S,H)| + |MP_e(S,H)|) = 2(|MP_o^S(S,H)| -
|MP_o^H(S,H)|).$$
\end{proof}

\bigskip

Note that lemmas \ref{S_and_B_are_the_same} and
\ref{LB_alternating_paths}, and corollary \ref{V_0=Q_LcupQ_B} are
the main properties of $A'$ that were mentioned above while
describing the idea of the proof, and are used further.

\bigskip

Now, we are going to construct a set of trails $A''$ mentioned
above, which is the key point of our proof.

First, let us prove two lemmas necessary for the construction of
$A''$.
\begin{lemma}\label{TL_in_endedges}
$T_L \subseteq E_1(D)$.
\end{lemma}
\begin{proof}
First let us proof that $T_L \subseteq E_1(A')$. suppose for
contradiction that there exists an edge $e = (u,v) \in T_L$, such
that $e$ is not an end-edge of any trail from $A'$. Since the set of
end-vertices of all trails from $A'$ is $Q_L \cup Q_B$ (corollary
\ref{V_0=Q_LcupQ_B}), there are trails $T_1, T_2 \in A'$ such that
$u$ and $v$ are the end-vertices of $T_1$ and $T_2$, respectively
($T_1$ and $T_2$ may coincide). As $A' \subseteq T_o^{L(F)}(L(F),
B(G,F))$, $e$ is adjacent to two edges from $L(F)$, which is
impossible because $e$ itself belongs to $L(F)$. Thus, $e$ is an
end-edge of some trail $T$ from $A'$.

Since $\overline{Z_{Q_U}} \subseteq H' \backslash (S \cup S')$ and
$e \in T_L \subseteq S' \backslash H'$, implies $T \in D$.
\end{proof}

\begin{lemma}\label{no_A'paths_share_the_same_edge}
No edge lies on two different trails from $A'$.
\end{lemma}
\begin{proof}
Clearly, the statement of the lemma is true for the following set of
paths: $MP(S,H) \cup \overline{Z_{Q_U}}.$ Let us prove that when
$MP(S,H)$ is transformed to $D$ it still remains true. During the
transformation of $MP(S,H)$ to $D$, the only edges we add are edges
$e''$ in the case \ref{case_einH_e'inB}, which belong to $H'
\backslash S$ (figure \ref{fig_trans1_1}). Clearly, each such edge
$e''$ is attached to only one end of one path from $MP(S,H)$, thus
lies on only one trail from $D$. Furthermore, due to lemma
\ref{H_in_U_is_TU} the edge $e$ shown in the figure
\ref{fig_trans1_1} belongs to $T_U$. Thus $e''$ does not belong to
$S'$, i.e. $e'' \in H' \backslash (S \cup S')$. Moreover, $e'' \in
Z_{Q_U}$ as the $2$-vertex incident to $e''$ belongs to $Q_U$ since
is incident to $e \in T_U$. Thus, $e'' \notin \overline{Z_{Q_U}}$,
and the proof of the lemma is complete.
\end{proof}

\begin{figure}[h]
\begin{center}
\includegraphics{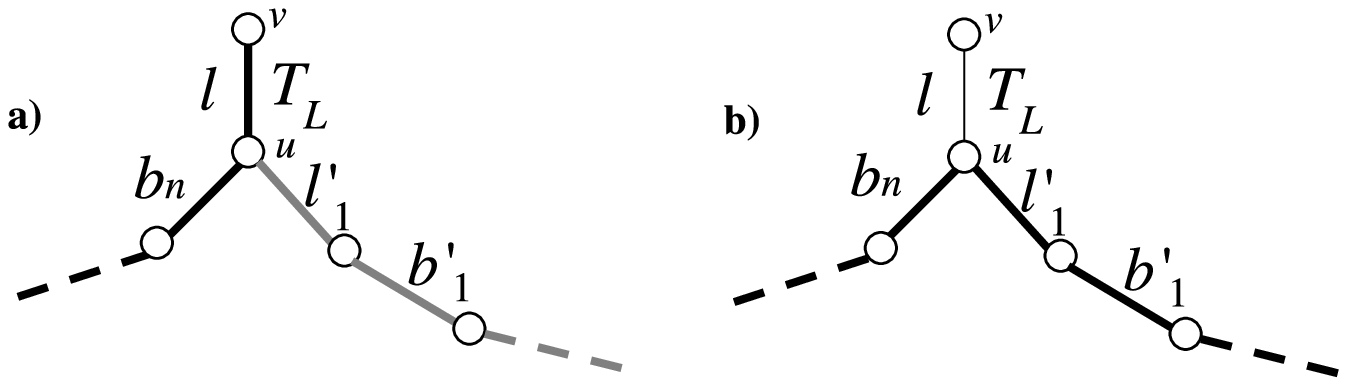}\\
\caption{}\label{fig_trans2}
\end{center}
\end{figure}

Let us introduce an operation called $JOIN$, using which we ``get
rid of" edges from $T_L$ ``preserving" the main properties obtained
for $A'$. Assume that $l = (u,v) \in T_L$, and $T_1 : l_1 b_1 l_2
... l_n b_n l$ and $T_2 : l_1' b_1' l_2' ... l_m' b_m' l_{m+1}'$ are
trails from $T_o^{L(F)}(L(F),B(G,F))$ such that $E(T_1) \cap E(T_2)
= \emptyset$, $v$ is the last vertex of $T_1$ incident to $l$ and
$u$ is the starting vertex of $T_2$ incident to $l_1'$. In this case
we say that $(T_1,T_2)$ is a $T_L$-adjacent pair of trails
corresponding to $l$ (figure \ref{fig_trans2}a).

If $(T_1, T_2)$ is a $T_L$-adjacent pair ($T_1, T_2 \in
T_o^{L(F)}(L(F),B(G,F))$) such that $T_1$ and $T_2$ do not coincide
then the $JOIN$ of $(T_1, T_2)$ is a trail defined as follows
(figure \ref{fig_trans2}b):
$$T = l_1 b_1 l_2 ... l_n b_n l_1' b_1' ... l_m' b_m' l_{m+1}'.$$
As $T_1$ and $T_2$ do not have common edges, this definition is
correct, i.e. $T$ is really a trail. Moreover, $T$ is not a cycle,
as $T_1$ and $T_2$ do not coincide ($T_1$ is not the reverse of
$T_2$). Thus, $T \in T_o^{L(F)}(L(F),B(G,F))$.

Now assume that $W \subseteq T_o^{L(F)}(L(F),B(G,F))$ such that no
two trails in it share a common edge. We say that the set $W'$ is a
$T_L$-reduction of $W$ if $W'$ is obtained from $W$ by removing
arbitrarily chosen $T_L$-adjacent trails $T_1, T_2 \in W$
corresponding to an edge $l=(u,v) \in T_L$ and, if $T_1$ and $T_2$
do not coincide, adding their $JOIN$. The following lemma is
obviously true for $W'$:
\begin{lemma}\label{W'_properties}\
\renewcommand{\labelenumi}{(\arabic{enumi})}
\begin{enumerate}
\item \label{W'_in_T_o} $W' \subseteq T_o^{L(F)}(L(F),B(G,F))$;
\item \label{preserve} no two trails in $W'$ share common edge;
\item \label{no_new_edge} no edge is added to $W'$, formally: $\displaystyle \bigcup_{T' \in W'}E(T') \subseteq \bigcup_{T \in W}E(T)$;
\item \label{end_edges_decrement} $E_1(W') = E_1(W) \backslash \{l\}$ and $V_0(W') = V_0(W) \backslash \{u, v\}$.
\end{enumerate}
\end{lemma}

Due to lemma \ref{TL_in_endedges}, for each edge from $T_L$ there is
a trail $T_1 : l_1 b_1 l_2 ... l_n b_n l$, $T_1 \in A'$. Let $l =
(u,v)$ and $v$ is the last vertex of $T_1$. Similarly, as $u \in
Q_L$, there is a trail $T_2 : l_1' b_1' l_2' ... l_m' b_m' l_{m+1}$,
$T_2 \in A'$, such that $u$ is the starting vertex of $T_2$ and is
incident to $l_1'$ (corollary \ref{V_0=Q_LcupQ_B}). Thus,
$(T_1,T_2)$ is a $T_L$-adjacent pair corresponding to $l$, for each
$l \in T_L$. As no two trails from $A'$ share a common edge (lemma
\ref{no_A'paths_share_the_same_edge}), $A'$ is applicable for
$T_L$-reduction operation. Consider a sequence of sets of trails
$A'=A_0, A_1,..., A_k, ...$, such that $A_i$ is a $T_L$-reduction of
$A_{i-1}$, for each $i = 1,2,...k$. Note that $k \leq |T_L|$ as
$T_L$-reduction operation can be applied not more than $|T_L|$
times. Furthermore, consider such a sequence $A'=A_0, A_1,..., A_k$
having maximum length. Due to lemma \ref{TL_in_endedges} and
statement (\ref{end_edges_decrement}) of lemma \ref{W'_properties},
$k = |T_L|$.

Set $A'' = A_k$.
\begin{lemma}\label{A''_properties}\
\renewcommand{\labelenumi}{(\arabic{enumi})}
\begin{enumerate}
\item \label{A''inT_o} $A'' \subseteq T_o^{L(F)}(L(F), B(G,F))$;
\item \label{S_and_B_are_the_same_in_A''} If an edge $e$ lies on a
trail from $A''$  and $e \in B(G,F) \cup S$ then $e \in B(G,F) \cap
S$;
\item \label{V_0=Q_B} $V_0(A'') = Q_B$.
\end{enumerate}
\end{lemma}
\begin{proof}
Statement (\ref{A''inT_o}) immediately follows from statement
(\ref{W'_in_T_o}) of lemma \ref{W'_properties}. Statement
(\ref{S_and_B_are_the_same_in_A''}) is implied from lemma
\ref{S_and_B_are_the_same} and statement (\ref{no_new_edge}) of
lemma \ref{W'_properties}. As $k = |T_L|$, all edges from $T_L$ are
``remove'' from trails of $A'$, i.e. no edge from $T_L$ lies on a
trail from $A''$. Thus no end-vertex of a trail from $A''$ belongs
to $Q_L$. As $V_0(A') = Q_L \cup Q_B$ (lemma \ref{V_0=Q_LcupQ_B}),
we get statement (\ref{V_0=Q_B}).
\end{proof}

The following is the main lemma of the Sufficiency part.
\begin{lemma}\label{LB_alt_cycles}
Any edge from $T_B$ lies on a cycle $C \in C_e(L(F), B(G,F))$ such
that $C_B$ is bipartite.
\end{lemma}
\begin{proof}
Due to statement (\ref{V_0=Q_B}) of lemma \ref{A''_properties}, each
vertex of $Q_B$ is an end-vertex of a path from $A''$. On the other
hand, it is incident to an edge from $T_B$. Furthermore, every
end-vertex of any path from $A''$ belongs to $Q_B$. Hence, every
edge of $T_B$ lies on a ({\em not necessarily simple}) cycle $C :
t_1, T_1, ..., t_r, T_r$ ($r \geq 1$), where $t_i \in T_B \subseteq
S', T_i \in A'', i = 1, 2, ..., r$ (figure \ref{fig_LB_alt_cycle}).

\begin{figure}[h]
\begin{center}
\includegraphics{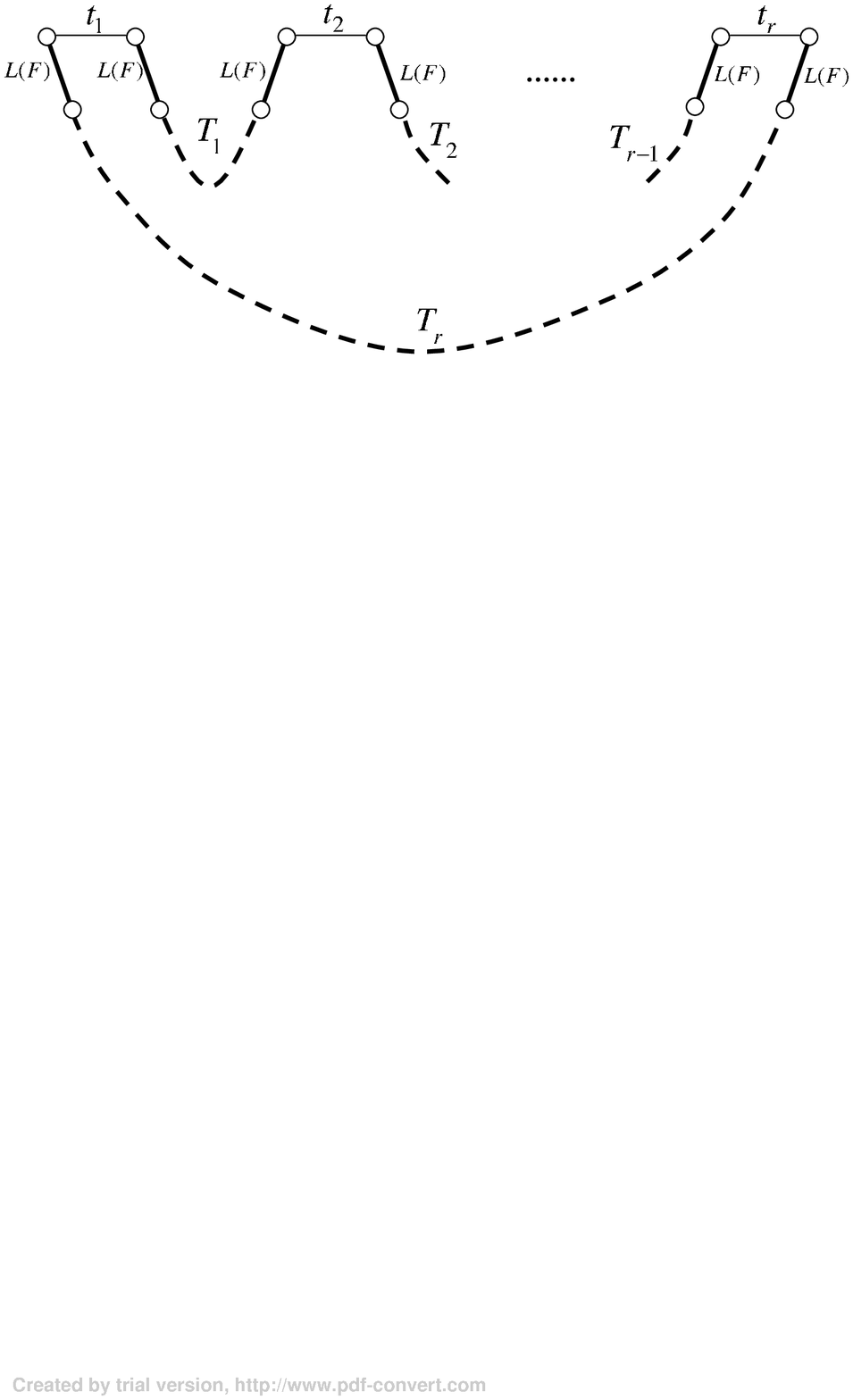}\\
\caption{}\label{fig_LB_alt_cycle}
\end{center}
\end{figure}

Due to statement (\ref{A''inT_o}) of lemma \ref{A''_properties} $C$
is an $L(F)$-$B(G,F)$ alternating cycle, i.e. $C \in C_e(L(F),
B(G,F))$. The construction of $C$ and statement
(\ref{S_and_B_are_the_same_in_A''}) of lemma \ref{A''_properties}
imply that $E(C_{B(G,F)}) \subseteq S \cup S'$ since $t_i \in S'$
and $E((T_i)_{B(G,F)}) \subseteq S$, $i = 1, 2, ..., r$. Therefore,
the graph $C_{B(G,F)}$ cannot contain an odd cycle, i.e. is
bipartite.
\end{proof}

\bigskip

Let us note that all lemmas and corollaries above in this subsection
are proved on the assumption that $G$ is an $S$-graph with spanning
$S$-forest $F$ satisfying the conditions (a) of the theorem and the
assumption \ref{assumption_no_delta} holds.

\bigskip

The following lemma is proved without any of the assumptions made
above.
\begin{lemma} \label{no_delta}
If $G'$ is an $S$-graph with spanning $S$-forest $F'$ satisfying the
condition (b) of the theorem then there is a pair $(S,S') \in
M_2(G')$ such that $(S \cup S') \cap \Delta(G', F') = \emptyset$.
\end{lemma}
\begin{proof}
Let $(S, S')$ be a pair from $M_2(G')$ such that $|(S \cup S') \cap
\Delta(G', F')|$ is minimum. Assume for contradiction that $|(S \cup
S') \cap \Delta(G', F')| \neq 0$. Let $e = (u, v) \in (S \cup S')
\cap \Delta(G, F)$, where $u$ is the $1$-vertex incident to $e$ and
$v$ is its base. Let $w$ be the $2$-vertex adjacent to $u$ and $v$.
The condition (b) of the theorem implies that $d_{G' \backslash
F'}(w) = 0$. Without loss of generality, we may assume that $e \in
S$. Let $e'$ be one of the two edges incident to $w$, which does not
belong to $S'$. Define a matching $S_1$ as follows:
$$S_1 \doteq (S \backslash \{ e \}) \cup \{e'\}.$$
Note that $(S_1, S') \in M_2(G')$ and $|(S_1 \cup S') \cap
\Delta(G', F')| < |(S \cup S') \cap \Delta(G', F')|$, which
contradicts the choice of $(S, S')$.
\end{proof}

\bigskip

Now assume that $G$ is an $S$-graph with spanning $S$-forest $F$
satisfying the conditions (a) and (b) of the theorem (do not assume
that the assumption \ref{assumption_no_delta} holds). Assume also
that $\frac{\beta(G)}{\alpha(G)} \neq \frac{5}{4}$, and therefore
$M_2(G) \cap M_2(F) = \emptyset$ (property
\ref{S-forest_lambda=2alpha=8k}). Thus, $M_2(G) \backslash M_2(F) =
M_2(G)$. This, together with lemma \ref{no_delta}, implies that
assumption \ref{assumption_no_delta} holds for $G$. Thus, all of the
lemmas and corollaries proved in this section are true for $G$.

\begin{lemma} \label{MP^S>MP^H}
$|MP_o^S(S,H)| > |MP_o^H(S,H)|$.
\end{lemma}
\begin{proof}
Corollary \ref{fivefourthinequality} and property
\ref{S-forest_lambda=2alpha=8k} imply that
$\frac{\beta(G)}{\alpha(G)} < \frac{5}{4} =
\frac{\beta(F)}{\alpha(F)}$. As $\beta(G) = \beta(F)$ (property
\ref{S-graph_beta=5k}), $\alpha(G) > \alpha(F)$, which means that
$|S| > |H|$. Thus, due to property \ref{cardinalitydiff}, we get:
$$|MP_o^S(S,H)| > |MP_o^H(S,H)|.$$
\end{proof}

Lemmas \ref{2-3} and \ref{MP^S>MP^H} together imply
\begin{corollary} \label{at_least_one_2-2}
There is at least one $2$-$2$ edge in $T_B$.
\end{corollary}

Let $e$ be one of $2$-$2$ edges from $T_B$. Due to lemma
\ref{LB_alt_cycles}, $e$ lies on a cycle $C \in C_e(L(F), B(G,F))$
such that $C_B$ is bipartite. Hence,

\begin{corollary} \label{does_not_satisfy_c}
The graph $G$ does not satisfy the condition (c) of the theorem.
\end{corollary}

\bigskip

Let us not that corollary \ref{does_not_satisfy_c} is proven on the
assumption that $G$ is an $S$-graph with spanning $S$-forest $F$
satisfying the conditions (a) and (b) of the theorem, and
$\frac{\beta(G)}{\alpha(G)} \neq \frac{5}{4}$. Clearly, this is
equivalent to the following:

\begin{statement}\label{sufficiency}
If $G$ is an $S$-graph with spanning $S$-forest $F$ satisfying the
conditions (a), (b) and (c) of the theorem then
$\frac{\beta(G)}{\alpha(G)} = \frac{5}{4}$.
\end{statement}

\subsection{Remarks}

\begin{remark}
Statements (\ref{necessity}) and (\ref{sufficiency}) imply that the
theorem can be reformulated as follows:

\textit{For a graph $G$ the equality
$\frac{\beta(G)}{\alpha(G)}=\frac{5}{4}$ holds, if and only if $G$
is an $S$-graph, \textbf{any} spanning $S$-forest of which satisfies
the conditions (a), (b) and (c).}
\end{remark}

\begin{remark}
Due to property \ref{2_and_3_are_equal}, the condition (c) of the
theorem can be changed to the following:

\textit{For every $L(F)$-$B(G,F)$ alternating even cycle $C$ of $G$
containing a $2$-$2$ or $3$-$3$ edge, the graph $C_{B(G,F)}$ is not
bipartite.}
\end{remark}

\vspace*{2cm}

\begin{acknowledgements}
\begin{center}
\end{center}
I would like to express my sincere gratitude to my supervisor Dr.
Vahan Mkrtchyan, for stating the problem and giving an opportunity
to perform a research in such an interesting area as Matching theory
is, for teaching me how to write articles and helping me to cope
with the current one, for directing and correcting my chaotic ideas,
for encouraging me, and finally, for his patience.

During the research process I also collaborated with my friend and
colleague Vahe Musoyan, who I want to thank for his great investment
in this work, for helping me to develop the main idea of
 the theorem, and for numerous of counter-examples that corrected my
statements as well as rejected some of my hypotheses.
\end{acknowledgements}

\newpage

\begin{center}

\end{center}

\end{document}